\documentclass[prd,preprintnumbers,longbibliography,
superscriptaddress,tightenlines,twocolumn,nofootinbib
]{revtex4-1}

\usepackage[utf8]{inputenc}
\usepackage{slashed}
\usepackage{amsmath}
\usepackage{amsfonts}
\usepackage{dsfont}
\usepackage{xcolor}
\usepackage{tikz}
\usepackage{tikz-feynman}
\usepackage{hyperref}
\usepackage{color}
\usepackage{orcidlink}
\usepackage{xspace}
\usepackage{placeins}

\usepackage{ulem}

\usetikzlibrary{decorations.shapes}
\usetikzlibrary{decorations.shapes,shapes.geometric} 

\graphicspath{{./Figures/}}

\newcommand{\dd}{\mathrm{d}}

\newcommand{\Neg}{\mathcal{N}}
\newcommand{\logNeg}{\mathcal{\epsilon}_\Neg}

\newcommand{\Negf}{\Neg^{2f}_{\alpha\beta}}
\newcommand{\logNegf}{\mathcal{\epsilon}_{\Neg_{2f}}}

\tikzset{decorate sep/.style 2 args=
{decorate,decoration={shape backgrounds,shape=circle,shape size=#1,shape sep=#2}}}
\tikzset{decorate sepbis/.style 2 args=
{decorate,decoration={shape backgrounds,shape=rectangle,shape size=#1,shape sep=#2}}}
\tikzset{decorate septris/.style 2 args=
{decorate,decoration={shape backgrounds,shape=diamond,shape size=#1,shape sep=#2}}}

\begin{document}

\title{Two-fermion negativity and confinement in the Schwinger model}

\author{Adrien Florio \,\orcidlink{0000-0002-7276-4515}}\email{aflorio@bnl.gov}
\affiliation{Department of Physics, Brookhaven National Laboratory, Upton, NY 11973}

\date{}

\begin{abstract}
We consider the fermionic (logarithmic) negativity between two fermionic modes in the Schwinger model. Recent results pointed out that fermionic systems can exhibit stronger entanglement than bosonic systems, exhibiting a negativity that decays only algebraically. The Schwinger model is described by fermionic excitations at short distances, while its asymptotic spectrum is the one of a bosonic theory. We show that the two-mode negativity detects this confining, fermion-to-boson transition, shifting from an algebraic decay to an exponential decay at distances of the order of the de Broglie wavelength of the first excited state. We derive analytical expressions in the massless Schwinger model and confront them with tensor network simulations. We also perform tensor network simulations in the massive model, which is not solvable analytically, and close to the Ising quantum critical point of the Schwinger model, where we show that the negativity behaves as its bosonic counterpart.
\end{abstract}

\date{\today}

\maketitle

\textit{Introduction.} Entanglement is a defining property of the quantum world. It is key to understand its structural impact on quantum many-body systems and quantum field theories. This was realized in the early days of modern quantum field theory \cite{Reeh:1961ujh}. It has led to a plethora of works, starting from the conjecture of the area law for the entanglement entropy \cite{Srednicki:1993im}, computation of the entanglement entropy in CFT \cite{Calabrese:2004eu}, and in holography \cite{Ryu:2006bv}. More than their intrinsic interests, these studies have led to reconsidering and gaining insights into old field theoretic questions and considering new problems. Examples range from understanding high-energy scattering in terms of entanglement \cite{Kharzeev:2017qzs,Kharzeev:2021yyf,Gong:2021bcp,deJong:2021wsd,Florio:2023dke,Belyansky:2023rgh,Barata:2023jgd,Lee:2023urk}, entanglement in pair production \cite{Florio:2021xvj,Dunne:2022zlx,Grieninger:2023ehb,Grieninger:2023pyb},  the impact of entanglement structures on the systematic construction of effective field theories \cite{Beane:2018oxh,Klco:2020rga,Klco:2021biu,Klco:2021cxq}, links to thermalization \cite{Kharzeev:2005iz,Berges:2017hne,Zhou:2021kdl,Mueller:2021gxd,Ebner:2023ixq,Yao:2023pht, Grieninger:2023ufa}, and to partial and dynamical symmetry breaking and restoration \cite{Ares:2022koq,Ares:2023kcz,Ares:2023ggj}, see also \cite{Nishioka:2018khk, Klco:2021lap, Bauer:2023qgm} for reviews on the interface of quantum information science and field theory.

To fully apprehend entanglement, the study of mixed states is crucial. From a many-body perspective, it allows, for instance, to probe entanglement between subsystems. In this case, entanglement measures are not unique. The (logarithmic) negativity is a useful one as it provides an upper bound on the amount of \textit{distillable} entanglement, i.e. the amount of entanglement that can be recovered using local observables \cite{Peres:1996dw,Horodecki:1997vt,Simon:1999lfr,Vidal:2002zz, Plenio:2005cwa}. Its behavior has been extensively studied in CFTs \cite{Calabrese:2012ew,Calabrese:2012nk,Murciano:2021djk,Murciano:2022vhe} and was also explored in free scalar field theories \cite{Klco:2020rga,Klco:2021biu,Klco:2021cxq}. Recently, Ref.~\cite{Parez:2023uxu} pointed out the interesting fact that fermionic degrees of freedom can carry more entanglement than bosonic degrees of freedom. In particular, they exhibited systems where the fermionic negativity decays only as a power law. This is in contrast to bosonic systems, as even close to critical points, the logarithmic negativity decays faster than any power \cite{Marcovitch:2008sxc, Calabrese:2012ew}. Having fermionic excitations potentially more strongly entangled than bosonic excitations is interesting in relation to confining theories, where the spectrum of the theory is not made out of trivial excitations of the microscopic particles. It is suggestive that when a fermionic theory gives rise only to bosonic excitations, the spatial distribution of negativity should probe the scale at which the theory confines. As a result, an intriguing problem to investigate is the relationship between entanglement measures and confinement.  

This work explores these ideas within the Schwinger model, quantum electrodynamics in $1+1$ dimensions:
\begin{equation}
    H = \int\dd x \left[
        \overline{\psi}\left(\gamma^1( - i \partial_1
      + A_1) +m\right)\psi \right
    ] + \tfrac{g^2}{2} E^2  \ ,
\label{eq:H:continuum}
\end{equation}
with $A_1$  the gauge potential, $E$ is the canonically conjugate
operator for the electric field and $\psi^T=(\psi_1\  \psi_2)$ is a two component spinor. Gauss's constraint supplements it,  
\begin{equation}
    \partial_1 E = \overline\psi \gamma^0 \psi \ \equiv \psi^\dagger \psi \ ,
\end{equation}
with $\bar \psi \equiv \psi^\dagger \gamma^0$. 
It is the manifestation of the remaining ``gauge invariance" of the Hamiltonian formalism.
We express the $\gamma$ matrices in terms of Pauli matrices as
\begin{equation}
   \gamma^0=\sigma_z\ ,\quad
   \gamma^1=-i \sigma_y \ , \quad
   \gamma^5 =\gamma^0\gamma^1=-\sigma_x \ .
\label{eq:gamma}
\end{equation}
It is an ideal playground to test these ideas as it is known to be dual to a bosonic scalar theory and that asymptotic excitations of the model, ``mesons", are bosons (they are solitons made out of the scalar field) \cite{Coleman:1975pw,Coleman:1976uz}. In other words, it behaves as a fermionic theory only over the shortest scales and transitions to a bosonic behavior at larger scales. Moreover, when the mass parameter $m$ vanishes, the bosonic theory is simply a \textit{massive} free scalar field with mass $m_b=g/\sqrt{\pi}$ \cite{Schwinger:1962tp}. In this case, the theory is analytically solvable.

\begin{figure}
    \centering
	\includegraphics{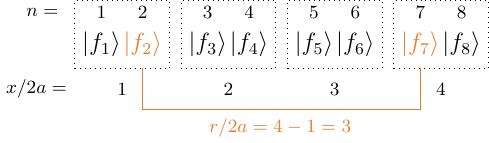}
\caption{Staggered set-up in the Schwinger model. Physical sites are made out of two lattice sites. The difference from a spinful ``electron" is that gauge fields, before being integrated out, also link fermions within a physical site. We consider here the negativity between even/odd modes; the same analysis can be repeated for the odd/odd or even/even correlations.}
\label{fig:setup}
\end{figure}

We will study a standard fermionic chain that approaches the Schwinger model in the continuum. It can be constructed by introducing ``staggered fermions" \cite{Kogut:1974ag, Susskind:1976jm}
\begin{equation}
    \chi_{2j} = \sqrt{2a} \psi_1(x_j), \ \chi_{2j+1} = \sqrt{2a} \psi_2(x_j) \ .
    \label{eq:staggeredfermions}
\end{equation}
As depicted in Fig.~\ref{fig:setup}, lattice sites are labeled by an integer $n=1,\dots,2N$. A physical site is made out of two lattice sites. The upper component of the continuum spinors lives on odd sites, while the down component lives on even sites.  Physical distances are measured in units of $1/2a$ with $a$ the lattice spacing. We further impose open-boundary conditions on the fermions $\chi_0=\chi_{2N+1}=0$. 
In this setting, the remaining gauge freedom allows to set the gauge potential to zero and express the electric field only in terms of fermionic operators by solving Gauss law, see for instance \cite{Ikeda:2020agk} for explicit expressions. The resulting Hamiltonian reads
\begin{align}
H_S^L &= -\frac{i}{2a}\sum_{n=1}^{N-1}
\big[\chi^\dag_{n}\chi_{n+1}-\chi^\dag_{n+1}\chi_{n}\big]+ m\sum_{n=1}^{N} (-1)^n \chi^\dag_n\chi_n 
\nonumber\\
&+\frac{ag^2}{2}\sum_{n=1}^{N-1}\left(\sum_{i=1}^n Q_i\right)^2\ , 
\end{align}
with $Q_i = n_i \left(1-(-1)^i\right)$ the lattice charge density operator and $n_i=\chi^\dagger_i \chi_i$ the fermion number operator. 

We will investigate the two-fermion (logarithmic) negativity. After defining it and studying its behavior in the continuum, we show that it indeed transitions from an algebraic decay at small distances to an exponential decay at large distances. Moreover, the transition happens precisely at the de Broglie wavelength of the first bosonic excited state. We conclude by investigating the behavior of the logarithmic fermionic negativity at the quantum Ising critical point of the Schwinger model. We show that even when using a fermionic description, the bosonic expectation of the theory is met, and the negativity decays exponentially fast.

\textit{Two-fermion negativity and the continuum limit.}
We start from the simple realization that a tractable expression for the density matrix for two staggered modes can be written down explicitly in terms of the staggered fields' two- and four-point functions \cite{Javanmard:2018xte,Berthiere:2021nkv,Parez:2023uxu}. We consider the reduced density matrix between modes whose physical sites are separated by a physical distance $r$ and are equally distant from the center of the lattice, as depicted in Fig.~\ref{fig:setup}. Because of the staggering, the sites are not related by a reflection symmetry, and we need a generalization of \cite{Parez:2023uxu}. Consider the state $|\psi\rangle = \sum_{i_1,\dots,i_{2N}=0,1} c_{i_1,\dots,i_{2N}}|i_1,\dots,i_{2N}\rangle$. The reduced density matrix obtained after tracing all sites except $j$ and $k$ reads
\begin{align}
    \rho_{j\leftrightarrow k} &= \sum_{i_j',i_k',i_j,i_k=0,1} \Lambda_{i_j'i_k',i_ji_k} |i_j'i_k'\rangle\langle i_ji_k| \ ,\label{eq:defrho} \\ 
    \Lambda_{i_j'i_k',i_ji_k} &= \sum_{\substack{\{i_l\}=0,1 \\ l\neq j,k}} c_{i_1,\dots i_j,\dots i_k, \dots,i_{2N}}^* c_{i_1,\dots i_j',\dots i_k',\dots,i_{2N}} \ . \notag
\end{align}
The matrix elements $\Lambda_{i_j'i_k',i_ji_k}$ can be expressed in terms of the fermion fields' two- and four-point functions. We restrict ourselves to states with zero total charge $Q=0$. As a result, the density matrix takes the following form
\begin{equation}
     \rho_{j\leftrightarrow k} = \begin{pmatrix}
    \Lambda_{00,00} &0 & 0 & 0 \\
    0 & \Lambda_{01,01} & \Lambda_{01,10} &0 \\
    0 & \Lambda_{01,10}^* & \Lambda_{10,10} & 0 \\
    0& 0 & 0 & \Lambda_{11,11} 
    \end{pmatrix} . \label{eq:rhoexpl}
\end{equation} \ 
Direct computations give
\begin{align}
    &\langle n_j \rangle_\psi =\Lambda_{10,10}+\Lambda_{11,11} , \hspace{-0.2cm} && \ \ \langle n_k \rangle_\psi =\Lambda_{01,01}+\Lambda_{11,11} ,\notag\\
    &\langle n_j n_k \rangle_\psi = \Lambda_{11,11} , \hspace{-0.2cm} && \ \ \langle \chi_j^\dagger \chi_k \rangle_\psi = \Lambda_{01,10} , \label{eq:rhocomp}
\end{align}
using the shorthand notation $\langle\dots\rangle_\psi \equiv \langle\psi|\dots|\psi\rangle$. 
 Thus, using \eqref{eq:rhocomp} and $\mathrm{Tr}\left( \rho_{j\leftrightarrow k}\right)=1$ in the occupation number basis $\{|00\rangle,|01\rangle,|10\rangle, |11\rangle\}$ , we have, 
\begin{align}
    &\Lambda_{00,00}=1-\langle n_j \rangle_\psi - \langle  n_k \rangle_\psi + \langle n_j n_k \rangle_\psi \ , \hspace{-3cm}\notag&\\
    &\Lambda_{10,10} = \langle n_j \rangle_\psi -\langle n_j n_k \rangle_\psi \ , & \Lambda_{11,11} = \langle n_j n_k \rangle_\psi \ ,   \notag \\
    &\Lambda_{01,01} = \langle n_k \rangle_\psi -\langle n_j n_k \rangle_\psi \ , & \Lambda_{01,10} = \langle \chi_j^\dagger \chi_k \rangle_\psi  \ .\notag  
\end{align}
We need the fermionic partial transpose of \eqref{eq:rhoexpl} to compute the negativity. Following \cite{Shapourian:2016cqu}, it acts on our two-fermion reduced Hilbert space as 
\begin{align}
(|i_j'i_k'\rangle\langle i_ji_k|)^{T^f_j} &= (-1)^\alpha |i_j i_k'\rangle\langle i_j' i_k| \ ,  \\
\alpha &= \frac{i_j'(i_j'+2)}{2} + \frac{i_j(i_j+2)}{2} \notag\\
&+ i_ki_k' + i_j'i_k'+i_ji_k+(i_j+i_k)(i_j'+i_k') \notag \ .
\end{align}
It leads to 
\begin{equation}
     \rho_{j\leftrightarrow k}^{T^j_j} = \begin{pmatrix}
    \Lambda_{00,00} &0 & 0 & i\Lambda_{01,10}^* \\
    0 & \Lambda_{01,01} & 0  &0 \\
    0 & 0 & \Lambda_{10,10} & 0 \\
    i\Lambda_{01,10}& 0 & 0 & \Lambda_{11,11} 
    \end{pmatrix} , \label{eq:rhoexpl}
\end{equation}
with eigenvalues $\lambda_1,\lambda_2,\lambda_+, \lambda_-$ :
\begin{align}
    \lambda_1 &= \Lambda_{01,01} \ \ , \ \ \ \ \ \ \  \lambda_2 = \Lambda_{10,10}  \ ,\notag\\
     \lambda_{\pm} &= \frac12\big(1-\Lambda_{01,01}-\Lambda_{10,10} \ \\
     &\pm \sqrt{-4|\Lambda_{01,10}|^2+(-\Lambda_{00,00}+\Lambda_{11,11})^2)} \big) \notag \ .
\end{align}
 The eigenvalues  $\lambda_{1,2}$ are real and positive. We can take  $\lambda_{\pm} = \lambda_R \pm i \lambda_I$ with $\lambda_I\neq 0$ as when they are real, they are also positive, as shown in App.~\ref{app:details_neg}. Then,
\begin{align}
    \Negf &= |\lambda_\pm| - \lambda_R \label{eq:neg_expr}\\ 
    \logNegf &= \log(1+2(|\lambda_\pm| - \lambda_R)) \label{eq:logneg_expr} \ .
\end{align}

 These simple expressions allow us to investigate what happens in the continuum limit. We focus on the ground state of the Schwinger model. First, we recover translational invariance.  Then, $\langle n_x \rangle = 1/2$ as the vacuum state of the field theory corresponds to half-filling in terms of the staggered fermions.  For the same reason, the correlator that has a continuum limit is the connected two-point function of $\langle \delta n_j \delta n_k \rangle$ (equivalent to ``normal ordering" in the continuum theory). Taking $x_j$ and $x_k$ the physical points associated with $j$ and $k$, see Fig.~\ref{fig:setup}, we write its continuum limit $n_2(r)$, with $r=x_j-x_k$ (it can be expressed in terms of the vector charge and the chiral condensate, but it won't be used here). Lastly, $\langle \chi^\dagger_j \chi_k \rangle$ converges to appropriate components of the continuum,  full equal time correlator of the theory $D_{\alpha\beta}(r)$. In more detail, we have, depending on their parity, 
 \begin{equation}
     \begin{pmatrix}
        \langle \chi^\dagger_{2x_j} \chi_{2x_k} \rangle & \langle \chi^\dagger_{2x_j} \chi_{2x_k+1} \rangle \\
        \langle \chi^\dagger_{2x_j+1} \chi_{2x_k} \rangle & \langle \chi^\dagger_{2x_j+1} \chi_{2x_k+1}  \rangle
     \end{pmatrix}{\to}     \begin{pmatrix}
        D_{11}(r) & D_{12}(r) \\
        D_{21}(r) & D_{22}(r)
     \end{pmatrix} \ .
 \end{equation}
 This is how the fact that the Dirac propagator in the continuum is a two-by-two matrix manifests itself in the staggered formulation.
 
 Reinstating the correct factors of $a$ according to 
\eqref{eq:staggeredfermions}, we have 
\begin{align}
    &\Lambda_{00,00}, \Lambda_{11,11} \to 2a^2 n_2 + \frac14  \ , \ \\
    &\Lambda_{10,10},\Lambda_{01,01}  \to \frac14 -2a^2n_2 \ \  , \  \Lambda_{01,10} \to 2a^2 D_{\alpha \beta}  \ \  \\
    &\lambda_\pm \to \frac14 -2a^2n_2\pm 2ia^2 D_{\alpha\beta}  \ .
\end{align}
Note that, notwithstanding the continuum analysis, it matches the results of \cite{Parez:2023uxu} around half-filling. 

 Now, we are in a position to study the continuum behavior of the two-fermion negativity.
Plugging this in \eqref{eq:neg_expr}, we find 
\begin{align}
    \Negf/a^2 \to 8|D_{\alpha\beta}|^2 + O(a^2) \ . \label{eq:contneg}
\end{align}
The fact that the two-fermion negativity vanishes as $a^2$ can easily be explained. The quantity with a continuum limit without any rescaling by factors of $a$ is the one between two subsystems of a given physical length. As $a\to 0$, the number of modes contained in each subsystem grows as $1/a$. By insisting on computing the negativity between single modes, we need to pay a factor of $1/a^2$ in \eqref{eq:contneg}. The indices $\alpha,\beta$ reflect the fact you can consider the negativity between the same component modes or opposite component modes. Note that this relation to the propagator is reminiscent of the results of \cite{Malla:2023cvl}, where the Quantum Fisher Information, another measure of multipartite entanglement, was successfully probed using single particle propagators only. 

What is insightful is that the continuum limit of $\Negf/a^2$ is simply proportional to the square propagator between the two modes. It gives a new interpretation to the space-like part of the propagator in this kind of field theory: it measures the two-mode entanglement in the system. It also strengthens the speculation made in the introduction. The two-fermion negativity detects confinement, at least in one dimension. More generally, it is suggestive that entanglement measures can be used to probe confinement, exhibiting different behavior when probing asymptotically free degrees of freedom with respect to scattering states.

\def\xsh{0.9}
\def\ysh{0.5}
\def\xcor{0.2}
\def\xmid{0.25}

\def\yshup{0.3}
\def\yshdn{0.5}

%\section{Results}
%\label{sec:results}
\textit{Massless Schwinger Model.} We exemplify these ideas by studying first the massless Schwinger model. The continuum theory is dual to the theory of a massive free boson $\phi$ of mass $m_b=g/\sqrt{\pi}$ \cite{Schwinger:1962tp}. The fermionic propagator is known explicitly \cite{Casher:1973uf, Steele:1994gf}
\begin{align}
D_{11}(r) &= D_{22}(r) \equiv d_1(r) =  \frac{m_b}{4\pi} e^{G(r)+\gamma_E}, \\ 
D_{12}(r) &= D_{21}(r) \equiv d_2(r)  = \frac{m_b}{4\pi} e^{G(r) + K_0(m_br)  + \gamma_E} , \\
G(r) &= -m_b^2\int_0^\infty \frac{\dd k}{\sqrt{k^2+m_b^2}} \frac{\sin^2\left(\frac{kr}{2}\right)}{k^2} \\
&= -\frac{m_b r}{8}  G_{2,4}^{2,2}\left(\frac{m_b^2 r^2}{4}\left|
\begin{array}{c}
 \frac{1}{2},1 \\
 \frac{1}{2},\frac{1}{2},-\frac{1}{2},0 \\
\end{array}
\right.
\right) \ ,
\end{align}
with $r=x-y$, $K_0$ the $0^{th}$ Bessel function of the second kind and $G$ is the Meijer $G$-function. They both admit simple asymptotes
\begin{align}
    d_1(r)\sim \begin{cases} \frac{m_b}{4\pi}e^{\gamma_E} \ \ \ \ \ \ \ \ \ \ \ \ , \  \ r\to 0  \\ \frac{m_b}{4\pi}e^{\frac{1}{2}+\gamma_E-\frac{m_b\pi r}{4}}  \ , ~ \ r\to \infty\end{cases}  , \\
    d_2(r)\sim \begin{cases} \frac{1}{2\pi r}  \ \ \ \ \ \ \ \ \ \ \ \ \ \ \ \ , \  \ r\to 0 \\ \frac{m_b}{4\pi}e^{\frac{1}{2}+\gamma_E-\frac{m_b\pi r}{4}} \ , ~ \ r\to \infty\end{cases} \ ,
\end{align}
with $\gamma_E$ the Euler-Mascheroni constant. As was pointed out early on \cite{Casher:1973uf}, the fermionic propagator captures the transition from a free constituent fermionic theory at small distances to a massive bosonic free bosonic theory at large distances, where the propagator is screened. 

\begin{figure}
    \centering
\includegraphics[scale=0.95]{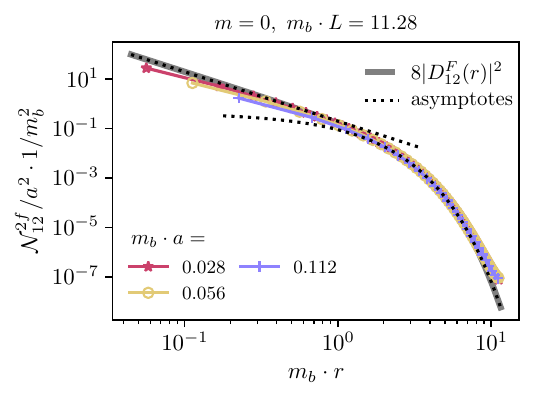}
    \caption{The deviations at small $r$ are lattice artifacts, as demonstrated by the apparent convergence to the analytical result as $a\to 0$. The deviations from the analytical prediction at large $r$ are due to boundary effects. The different lattice spacings correspond to lattices of $N=100,200,400$ staggered sites.}
    \label{fig:masslessneg}
\end{figure}

We now show that the two-fermion negativity defined from the lattice model indeed approaches the square propagator. To compute the negativity from the lattice model, we performed tensor network simulations using the ITensor library \cite{itensor, itensor-r0.3}. We compute the ground state using DMRG and evaluate the negativity with \eqref{eq:neg_expr}. To minimize finite-size effects, and for all simulations presented in this work, we use the improved lattice mass parameter of \cite{Dempsey:2022nys}. We show the even/odd negativity $\Neg^{2f}_{12}$ normalized by $a^2$ for a fixed coupling in Fig.~\ref{fig:masslessneg}. We also fix the physical volume such that finite-volume effects are negligible. The colored curves correspond to different lattice spacings and represent taking the continuum limit (with $N=100, 200$, and $400$ staggered sites). The grey curve is the analytical prediction \eqref{eq:contneg}. As claimed in the previous section, the negativity systematically approaches the norm square of the propagator. The transition from the algebraic fermionic behavior to the exponentially decreasing bosonic one is sharp and happens at the boson radius $1/m_b$. Note the deviations from the analytical prediction at a large radius are due to approaching the boundaries of our lattice (we use open-boundary conditions). It is a concrete example where the two sites' fermionic negativity transitions from an algebraic decay to an exponential one. This transition probes confinement.  We will see that it still happens in massive case which is not analytically solvable.  

%\label{sec:massless}

\begin{figure}
    \centering
    \includegraphics[scale=0.95]{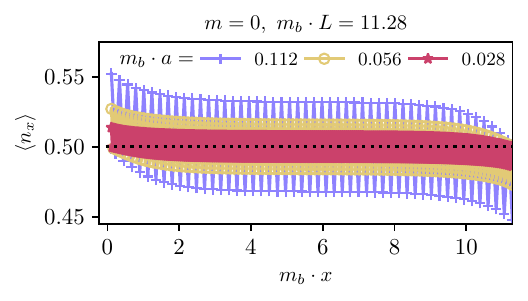}
    \caption{Number density as a function of distance. The continuum vacuum is at half-filling. Odd sites approach half-filling from above and even sites from below. }
    \label{fig:ndensity}
\end{figure}

The same analysis can, in principle, be repeated for the even/even or odd/odd case. In practice and because of the use of staggered fermions, it turns out to be more complicated. The cancellation between $\Lambda_{00,00}$ and $\Lambda_{11,11}$ is crucial to get a non-zero imaginary part and happens only when $\langle n_{j}\rangle +\langle n_k \rangle$ is sufficiently close to $1$. We show the typical behavior of $\langle n_x \rangle$ in Fig.~\ref{fig:ndensity}. Both odd and even site densities approach $1/2$ rather slowly in an alternating pattern. On the other hand, the sum of even and odd densities approaches one much faster as $a\to 0$. As a result, the even/even and odd/odd negativities are zero for the range of lattice spacings we studied and become non-zero only closer to the continuum. Consequently, we will restrict ourselves to the study of the even/odd negativity for the rest of this work.

\begin{figure}
    \centering
    \includegraphics[scale=0.95]{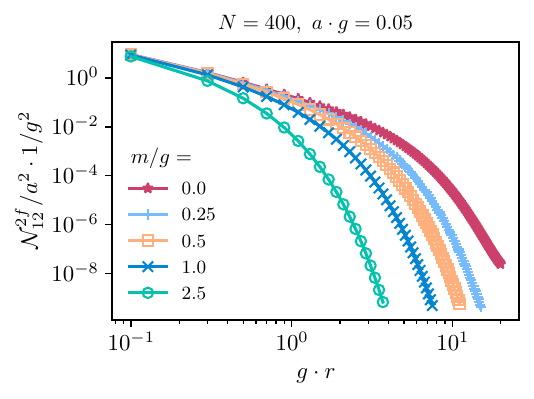}
    \caption{Two-fermion negativity as a function of distance, for different $m/g$ ratios. All curves transition from an algebraic to an exponential decay.}
    \label{fig:massiveneg}
\end{figure}

\begin{figure}
    \centering
    \includegraphics[scale=0.95]{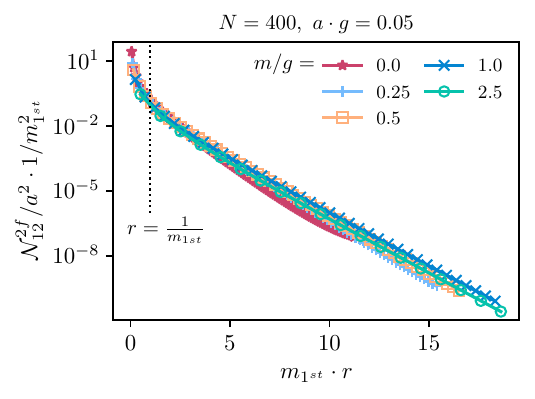}
    \caption{Two-fermion negativity as a function of distance, for different $m/g$ ratios, with the axis scaled by appropriate powers of the mass gap. The curves' approximate collapse indicates that the negativity's main dependence is on the mass gap. The transition from the algebraic fermionic decay to the bosonic exponential one happens at $r\sim 1/m_{1st}$, indicated by a dotted black line. }
    \label{fig:massivenegscaled}
\end{figure}

\textit{Massive Schwinger model.} We now move on to the massive Schwinger model, which is not known to be analytically solvable. It is dual to an interacting bosonic theory \cite{Coleman:1976uz}. We repeat our DMRG simulation for the same $g$ value as in the massless case for different mass parameters.  We show the resulting negativities in Fig~\ref{fig:massiveneg}. The transition from an algebraic behavior to an exponential decay is still clear. It happens at shorter distances for larger masses. We show the same data in Fig.~\ref{fig:massivenegscaled}, rescaling the $x$ and $y$ axis in terms of the mass gap. For the three smaller ratios of $m/g$, we used the values of \cite{Adam:2002pd}, which we checked against our own determination of the mass gap obtained by targeting the first excited state. For the larger mass, the leading prediction $m_{1st}^2 = m_b^2 + m^2$ is accurate at the percent level \cite{Adam:2002pd} and was used as an estimate for the gap. We observe a relatively good collapse, showing that the leading dependence of the negativity is the mass of the first excited state. In particular, the transition from an algebraic behavior to an exponential decay happens in all cases at $r\sim1/m_{1st}$, and thus is sensitive to the typical confinement length at which the theory transitions from being fermionic to bosonic.
%\label{sec:massive}

\begin{figure}
    \centering
    \includegraphics[scale=0.95]{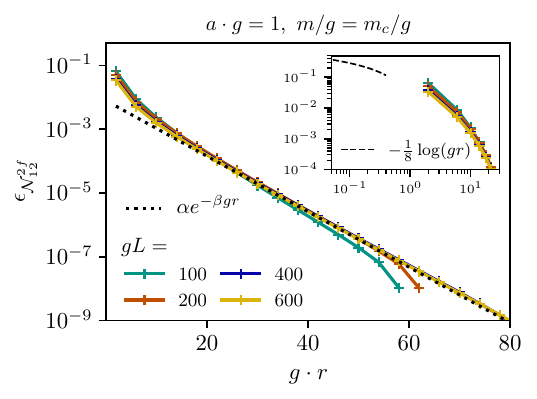}
    \caption{Two-fermion logarithmic negativity at the critical mass. At large distances, the fermionic negativity still decays exponentially, as expected for the $2D$ Ising CFT. The dashed line is the result of a fit to an exponential $\alpha e^{-\beta gr}$ with $\alpha=0.00786 \pm 0.00023$ and $\beta = 0.20051 \pm 0.00075$. Inset: while out-of-reach of the data presented here, the behavior of $\logNeg$ also seems consistent with the small $r$ prediction $\logNeg \sim -c/4\log(gr)$ with $c=1/2$ the central charge of the $2D$ Ising CFT. }
    \label{fig:criticalneg}
\end{figure}

\textit{Massive Schwinger model at criticality.} The Schwinger is critical for negative masses at the critical mass to coupling ratio $m_c/g\approx 0.3335$ \cite{Byrnes:2002gj,Byrnes:2002nv} and is in the $2D$ Ising universality class. Ref.~\cite{Parez:2023uxu} showed that fermionic CFTs can exhibit logarithmic negativity that decays algebraically, in contrast to bosonic CFTs where the logarithmic negativity is known to decay faster than any polynomials \cite{Marcovitch:2008sxc, Calabrese:2012ew}. An interesting feature of the massive Schwinger model is that it is a fermionic representation of a bosonic CFT, and it is thus of value to investigate the behavior of the fermionic logarithmic negativity in this model. Results are shown in Fig.~\ref{fig:criticalneg}. The system is tuned at its second-order critical point.  The two-fermion logarithmic negativity is plotted as a function of distance. The different curves correspond to different volumes. The dashed line is a fit to an exponential decay. In this case, the fermionic negativity is also exponentially suppressed at large distances; it behaves in the same way as the bosonic one. This matches expectations from universality, as the underlying CFT has no knowledge of fermionic operators. 

We found the small distance behavior to be out of reach of the specific numerical methods used in this work; we show the small $r$ behavior of our data in the inset. We also plot the universal Ising expectation for very small $r$. While far from the regime of applicability of this prediction, it appears that our data do not contradict this small distance behavior. This regime can be better studied by using dedicated tensor network simulations more appropriate to critical systems \cite{Vidal:2009cbi, Bridgeman:2016dhh}.
%\label{sec:critical}

\textit{Discussion.} We studied in this work the two-fermion (logarithmic) negativity of fermionic spin chains and its behavior in the continuum limit. In particular, we focused on the Schwinger model. While expressible purely as a fermionic chain, it has the interesting property of being fermionic only at short distances, while its asymptotic spectrum is described by a bosonic scalar theory. We showed that the negativity is sensitive to this ``confinement radius", at which it transitions from an algebraic decay to an exponential one.  We also showed that the fermionic negativity behaves as its bosonic counterpart at the Ising critical point of the massive Schwinger model, as universality would suggest.

These results are interesting from two different perspectives. First, it is an explicit example where the fermionic negativity exhibits a transition from a more entangling, in the sense of \cite{Parez:2023uxu}, fermionic behavior to a bosonic behavior at large distances, as would be expected by bosonization. It shows that the fermionic negativity captures the system's fundamental statistics and does not measure an artifact related to the representation of the system's degrees of freedom. Second, it further motivates us to think about how confinement relates to entanglement and how entanglement measures can be used to probe it. Moreover, the explicit relation between the two-fermion negativity and the full propagator of the theory is intriguing and invites reflection on how entanglement measures between physical subregions can be systematically related to field theoretical $n$-point functions.

This work opens other natural outlooks. A simple extension would be to consider how the two-mode negativity is affected when computed between two full Dirac fermions (four modes negativity in a staggered formulation or two-site negativity for Wilson fermions). Considering its behavior in the presence of localized states is also an interesting avenue.  More challenging, investigating the two-mode density matrix in the presence of dynamical gauge fields will be an essential step in generalizing these results to higher dimensions and more realistic theories. 

\textit{Acknowledgment.} The author thanks Jo\~ao Barata for interesting discussions, Erik Gustafson for speeding up his code, and Robert Szafron for insightful discussions on negativity. The author is also grateful to Dmitri Kharzeev, Robert Pisarski, Robert Szafron, and Raju Venugopalan for comments on a draft of this manuscript.
 The author is supported by the U.S. Department of Energy under contract 
DE-SC0012704 and by the U.S. 
Department of Energy, Office of Science, National Quantum 
Information Science Research Centers, Co-design Center
for Quantum Advantage (C$^2$QA), under contract number
DE-SC0012704.

\FloatBarrier
%\clearpage

\appendix

\section{Details on the derivation of the negativity}
\label{app:details_neg}
 
To compute the (logarithmic) negativity, we consider the sum of the absolute values of these eigenvalues. We start by noting that $\Lambda_{00,00}, \Lambda_{01,01}, \Lambda_{10,10}, \Lambda_{11,11} \geq 0$ as they are all sums of positive numbers by definition \eqref{eq:defrho} in the main text. This implies $\lambda_1,\lambda_2>0$. Now assume $\lambda_\pm$ are real. We have the following inequalities 
\begin{widetext}
    \begin{align}
     \lambda_{+} &= \frac12\left(1-\Lambda_{01,01}-\Lambda_{10,10}
     + \sqrt{-4|\Lambda_{01,10}|^2+(-\Lambda_{00,00}+\Lambda_{11,11})^2)} \right) \\
     &\geq \frac12\left(1-\Lambda_{01,01}-\Lambda_{10,10}\right) = \frac12\left(\Lambda_{00,00} + \Lambda_{11,11}  \right) \geq 0 \\
     \lambda_{-} &= \frac12\left(1-\Lambda_{01,01}-\Lambda_{10,10}
     - \sqrt{-4|\Lambda_{01,10}|^2+(-\Lambda_{00,00}+\Lambda_{11,11})^2)} \right) \\
     &\geq \frac12\left(1-\Lambda_{01,01}-\Lambda_{10,10}
     - |-\Lambda_{00,00}+\Lambda_{11,11}| \right) \\
     &= \begin{cases} \Lambda_{00,00} \geq 0 \ , \text{ if } \Lambda_{00,00}\leq\Lambda_{11,11}\\
     \Lambda_{11,11} \geq 0 \ , \text{ if } \Lambda_{00,00}\geq\Lambda_{11,11}
     \end{cases}  \ .
     \end{align}
\end{widetext}
In this case, the (logarithmic) negativity vanishes.

The other case to consider is when $\lambda_{\pm} = \lambda_R \pm i \lambda_I$ are complex conjugate to each other with $\lambda_I\neq 0$. We then have 
\begin{align}
    |\lambda_1| + |\lambda_2| + 2|\lambda_{\pm}| &= 1 - \lambda_+ -\lambda_-  +2|\lambda_{\pm}|  \\
    &= 1- 2(|\lambda_\pm| - \lambda_R)
\end{align}
and Eqs. \eqref{eq:neg_expr}-\eqref{eq:logneg_expr} trivially follow from the definition of the (logarithmic) negativity.

%\bibliography{main.bib, manual.bib}

\begin{thebibliography}{64}%
\makeatletter
\providecommand \@ifxundefined [1]{%
 \@ifx{#1\undefined}
}%
\providecommand \@ifnum [1]{%
 \ifnum #1\expandafter \@firstoftwo
 \else \expandafter \@secondoftwo
 \fi
}%
\providecommand \@ifx [1]{%
 \ifx #1\expandafter \@firstoftwo
 \else \expandafter \@secondoftwo
 \fi
}%
\providecommand \natexlab [1]{#1}%
\providecommand \enquote  [1]{``#1''}%
\providecommand \bibnamefont  [1]{#1}%
\providecommand \bibfnamefont [1]{#1}%
\providecommand \citenamefont [1]{#1}%
\providecommand \href@noop [0]{\@secondoftwo}%
\providecommand \href [0]{\begingroup \@sanitize@url \@href}%
\providecommand \@href[1]{\@@startlink{#1}\@@href}%
\providecommand \@@href[1]{\endgroup#1\@@endlink}%
\providecommand \@sanitize@url [0]{\catcode `\\12\catcode `\$12\catcode
  `\&12\catcode `\#12\catcode `\^12\catcode `\_12\catcode `\%12\relax}%
\providecommand \@@startlink[1]{}%
\providecommand \@@endlink[0]{}%
\providecommand \url  [0]{\begingroup\@sanitize@url \@url }%
\providecommand \@url [1]{\endgroup\@href {#1}{\urlprefix }}%
\providecommand \urlprefix  [0]{URL }%
\providecommand \Eprint [0]{\href }%
\providecommand \doibase [0]{http://dx.doi.org/}%
\providecommand \selectlanguage [0]{\@gobble}%
\providecommand \bibinfo  [0]{\@secondoftwo}%
\providecommand \bibfield  [0]{\@secondoftwo}%
\providecommand \translation [1]{[#1]}%
\providecommand \BibitemOpen [0]{}%
\providecommand \bibitemStop [0]{}%
\providecommand \bibitemNoStop [0]{.\EOS\space}%
\providecommand \EOS [0]{\spacefactor3000\relax}%
\providecommand \BibitemShut  [1]{\csname bibitem#1\endcsname}%
\let\auto@bib@innerbib\@empty
%</preamble>
\bibitem [{\citenamefont {Reeh}\ and\ \citenamefont
  {Schlieder}(1961)}]{Reeh:1961ujh}%
  \BibitemOpen
  \bibfield  {author} {\bibinfo {author} {\bibfnamefont {H.}~\bibnamefont
  {Reeh}}\ and\ \bibinfo {author} {\bibfnamefont {S.}~\bibnamefont
  {Schlieder}},\ }\bibfield  {title} {\enquote {\bibinfo {title} {{Bemerkungen
  zur unit\"ar\"aquivalenz von lorentzinvarianten feldern}},}\ }\href {\doibase
  10.1007/BF02787889} {\bibfield  {journal} {\bibinfo  {journal} {Nuovo Cim.}\
  }\textbf {\bibinfo {volume} {22}},\ \bibinfo {pages} {1051--1068} (\bibinfo
  {year} {1961})}\BibitemShut {NoStop}%
\bibitem [{\citenamefont {Srednicki}(1993)}]{Srednicki:1993im}%
  \BibitemOpen
  \bibfield  {author} {\bibinfo {author} {\bibfnamefont {Mark}\ \bibnamefont
  {Srednicki}},\ }\bibfield  {title} {\enquote {\bibinfo {title} {{Entropy and
  area}},}\ }\href {\doibase 10.1103/PhysRevLett.71.666} {\bibfield  {journal}
  {\bibinfo  {journal} {Phys. Rev. Lett.}\ }\textbf {\bibinfo {volume} {71}},\
  \bibinfo {pages} {666--669} (\bibinfo {year} {1993})},\ \Eprint
  {http://arxiv.org/abs/hep-th/9303048} {arXiv:hep-th/9303048} \BibitemShut
  {NoStop}%
\bibitem [{\citenamefont {Calabrese}\ and\ \citenamefont
  {Cardy}(2004)}]{Calabrese:2004eu}%
  \BibitemOpen
  \bibfield  {author} {\bibinfo {author} {\bibfnamefont {Pasquale}\
  \bibnamefont {Calabrese}}\ and\ \bibinfo {author} {\bibfnamefont {John~L.}\
  \bibnamefont {Cardy}},\ }\bibfield  {title} {\enquote {\bibinfo {title}
  {{Entanglement entropy and quantum field theory}},}\ }\href {\doibase
  10.1088/1742-5468/2004/06/P06002} {\bibfield  {journal} {\bibinfo  {journal}
  {J. Stat. Mech.}\ }\textbf {\bibinfo {volume} {0406}},\ \bibinfo {pages}
  {P06002} (\bibinfo {year} {2004})},\ \Eprint
  {http://arxiv.org/abs/hep-th/0405152} {arXiv:hep-th/0405152} \BibitemShut
  {NoStop}%
\bibitem [{\citenamefont {Ryu}\ and\ \citenamefont
  {Takayanagi}(2006)}]{Ryu:2006bv}%
  \BibitemOpen
  \bibfield  {author} {\bibinfo {author} {\bibfnamefont {Shinsei}\ \bibnamefont
  {Ryu}}\ and\ \bibinfo {author} {\bibfnamefont {Tadashi}\ \bibnamefont
  {Takayanagi}},\ }\bibfield  {title} {\enquote {\bibinfo {title} {{Holographic
  derivation of entanglement entropy from AdS/CFT}},}\ }\href {\doibase
  10.1103/PhysRevLett.96.181602} {\bibfield  {journal} {\bibinfo  {journal}
  {Phys. Rev. Lett.}\ }\textbf {\bibinfo {volume} {96}},\ \bibinfo {pages}
  {181602} (\bibinfo {year} {2006})},\ \Eprint
  {http://arxiv.org/abs/hep-th/0603001} {arXiv:hep-th/0603001} \BibitemShut
  {NoStop}%
\bibitem [{\citenamefont {Kharzeev}\ and\ \citenamefont
  {Levin}(2017)}]{Kharzeev:2017qzs}%
  \BibitemOpen
  \bibfield  {author} {\bibinfo {author} {\bibfnamefont {Dmitri~E.}\
  \bibnamefont {Kharzeev}}\ and\ \bibinfo {author} {\bibfnamefont {Eugene~M.}\
  \bibnamefont {Levin}},\ }\bibfield  {title} {\enquote {\bibinfo {title}
  {{Deep inelastic scattering as a probe of entanglement}},}\ }\href {\doibase
  10.1103/PhysRevD.95.114008} {\bibfield  {journal} {\bibinfo  {journal} {Phys.
  Rev. D}\ }\textbf {\bibinfo {volume} {95}},\ \bibinfo {pages} {114008}
  (\bibinfo {year} {2017})},\ \Eprint {http://arxiv.org/abs/1702.03489}
  {arXiv:1702.03489 [hep-ph]} \BibitemShut {NoStop}%
\bibitem [{\citenamefont {Kharzeev}\ and\ \citenamefont
  {Levin}(2021)}]{Kharzeev:2021yyf}%
  \BibitemOpen
  \bibfield  {author} {\bibinfo {author} {\bibfnamefont {Dmitri~E.}\
  \bibnamefont {Kharzeev}}\ and\ \bibinfo {author} {\bibfnamefont {Eugene}\
  \bibnamefont {Levin}},\ }\bibfield  {title} {\enquote {\bibinfo {title}
  {{Deep inelastic scattering as a probe of entanglement: Confronting
  experimental data}},}\ }\href {\doibase 10.1103/PhysRevD.104.L031503}
  {\bibfield  {journal} {\bibinfo  {journal} {Phys. Rev. D}\ }\textbf {\bibinfo
  {volume} {104}},\ \bibinfo {pages} {L031503} (\bibinfo {year} {2021})},\
  \Eprint {http://arxiv.org/abs/2102.09773} {arXiv:2102.09773 [hep-ph]}
  \BibitemShut {NoStop}%
\bibitem [{\citenamefont {Gong}\ \emph {et~al.}(2022)\citenamefont {Gong},
  \citenamefont {Parida}, \citenamefont {Tu},\ and\ \citenamefont
  {Venugopalan}}]{Gong:2021bcp}%
  \BibitemOpen
  \bibfield  {author} {\bibinfo {author} {\bibfnamefont {Wenjie}\ \bibnamefont
  {Gong}}, \bibinfo {author} {\bibfnamefont {Ganesh}\ \bibnamefont {Parida}},
  \bibinfo {author} {\bibfnamefont {Zhoudunming}\ \bibnamefont {Tu}}, \ and\
  \bibinfo {author} {\bibfnamefont {Raju}\ \bibnamefont {Venugopalan}},\
  }\bibfield  {title} {\enquote {\bibinfo {title} {{Measurement of Bell-type
  inequalities and quantum entanglement from \ensuremath{\Lambda}-hyperon spin
  correlations at high energy colliders}},}\ }\href {\doibase
  10.1103/PhysRevD.106.L031501} {\bibfield  {journal} {\bibinfo  {journal}
  {Phys. Rev. D}\ }\textbf {\bibinfo {volume} {106}},\ \bibinfo {pages}
  {L031501} (\bibinfo {year} {2022})},\ \Eprint
  {http://arxiv.org/abs/2107.13007} {arXiv:2107.13007 [hep-ph]} \BibitemShut
  {NoStop}%
\bibitem [{\citenamefont {de~Jong}\ \emph {et~al.}(2022)\citenamefont
  {de~Jong}, \citenamefont {Lee}, \citenamefont {Mulligan}, \citenamefont
  {P\l{}osko\'n}, \citenamefont {Ringer},\ and\ \citenamefont
  {Yao}}]{deJong:2021wsd}%
  \BibitemOpen
  \bibfield  {author} {\bibinfo {author} {\bibfnamefont {Wibe~A.}\ \bibnamefont
  {de~Jong}}, \bibinfo {author} {\bibfnamefont {Kyle}\ \bibnamefont {Lee}},
  \bibinfo {author} {\bibfnamefont {James}\ \bibnamefont {Mulligan}}, \bibinfo
  {author} {\bibfnamefont {Mateusz}\ \bibnamefont {P\l{}osko\'n}}, \bibinfo
  {author} {\bibfnamefont {Felix}\ \bibnamefont {Ringer}}, \ and\ \bibinfo
  {author} {\bibfnamefont {Xiaojun}\ \bibnamefont {Yao}},\ }\bibfield  {title}
  {\enquote {\bibinfo {title} {{Quantum simulation of nonequilibrium dynamics
  and thermalization in the Schwinger model}},}\ }\href {\doibase
  10.1103/PhysRevD.106.054508} {\bibfield  {journal} {\bibinfo  {journal}
  {Phys. Rev. D}\ }\textbf {\bibinfo {volume} {106}},\ \bibinfo {pages}
  {054508} (\bibinfo {year} {2022})},\ \Eprint
  {http://arxiv.org/abs/2106.08394} {arXiv:2106.08394 [quant-ph]} \BibitemShut
  {NoStop}%
\bibitem [{\citenamefont {Florio}\ \emph {et~al.}(2023)\citenamefont {Florio},
  \citenamefont {Frenklakh}, \citenamefont {Ikeda}, \citenamefont {Kharzeev},
  \citenamefont {Korepin}, \citenamefont {Shi},\ and\ \citenamefont
  {Yu}}]{Florio:2023dke}%
  \BibitemOpen
  \bibfield  {author} {\bibinfo {author} {\bibfnamefont {Adrien}\ \bibnamefont
  {Florio}}, \bibinfo {author} {\bibfnamefont {David}\ \bibnamefont
  {Frenklakh}}, \bibinfo {author} {\bibfnamefont {Kazuki}\ \bibnamefont
  {Ikeda}}, \bibinfo {author} {\bibfnamefont {Dmitri}\ \bibnamefont
  {Kharzeev}}, \bibinfo {author} {\bibfnamefont {Vladimir}\ \bibnamefont
  {Korepin}}, \bibinfo {author} {\bibfnamefont {Shuzhe}\ \bibnamefont {Shi}}, \
  and\ \bibinfo {author} {\bibfnamefont {Kwangmin}\ \bibnamefont {Yu}},\
  }\bibfield  {title} {\enquote {\bibinfo {title} {{Real-Time Nonperturbative
  Dynamics of Jet Production in Schwinger Model: Quantum Entanglement and
  Vacuum Modification}},}\ }\href {\doibase 10.1103/PhysRevLett.131.021902}
  {\bibfield  {journal} {\bibinfo  {journal} {Phys. Rev. Lett.}\ }\textbf
  {\bibinfo {volume} {131}},\ \bibinfo {pages} {021902} (\bibinfo {year}
  {2023})},\ \Eprint {http://arxiv.org/abs/2301.11991} {arXiv:2301.11991
  [hep-ph]} \BibitemShut {NoStop}%
\bibitem [{\citenamefont {Belyansky}\ \emph {et~al.}(2023)\citenamefont
  {Belyansky}, \citenamefont {Whitsitt}, \citenamefont {Mueller}, \citenamefont
  {Fahimniya}, \citenamefont {Bennewitz}, \citenamefont {Davoudi},\ and\
  \citenamefont {Gorshkov}}]{Belyansky:2023rgh}%
  \BibitemOpen
  \bibfield  {author} {\bibinfo {author} {\bibfnamefont {Ron}\ \bibnamefont
  {Belyansky}}, \bibinfo {author} {\bibfnamefont {Seth}\ \bibnamefont
  {Whitsitt}}, \bibinfo {author} {\bibfnamefont {Niklas}\ \bibnamefont
  {Mueller}}, \bibinfo {author} {\bibfnamefont {Ali}\ \bibnamefont
  {Fahimniya}}, \bibinfo {author} {\bibfnamefont {Elizabeth~R.}\ \bibnamefont
  {Bennewitz}}, \bibinfo {author} {\bibfnamefont {Zohreh}\ \bibnamefont
  {Davoudi}}, \ and\ \bibinfo {author} {\bibfnamefont {Alexey~V.}\ \bibnamefont
  {Gorshkov}},\ }\bibfield  {title} {\enquote {\bibinfo {title} {{High-Energy
  Collision of Quarks and Hadrons in the Schwinger Model: From Tensor Networks
  to Circuit QED}},}\ }\href@noop {} {\  (\bibinfo {year} {2023})},\ \Eprint
  {http://arxiv.org/abs/2307.02522} {arXiv:2307.02522 [quant-ph]} \BibitemShut
  {NoStop}%
\bibitem [{\citenamefont {Barata}\ \emph {et~al.}(2023)\citenamefont {Barata},
  \citenamefont {Gong},\ and\ \citenamefont {Venugopalan}}]{Barata:2023jgd}%
  \BibitemOpen
  \bibfield  {author} {\bibinfo {author} {\bibfnamefont {Jo\~ao}\ \bibnamefont
  {Barata}}, \bibinfo {author} {\bibfnamefont {Wenjie}\ \bibnamefont {Gong}}, \
  and\ \bibinfo {author} {\bibfnamefont {Raju}\ \bibnamefont {Venugopalan}},\
  }\bibfield  {title} {\enquote {\bibinfo {title} {{Realtime dynamics of
  hyperon spin correlations from string fragmentation in a deformed four-flavor
  Schwinger model}},}\ }\href@noop {} {\  (\bibinfo {year} {2023})},\ \Eprint
  {http://arxiv.org/abs/2308.13596} {arXiv:2308.13596 [hep-ph]} \BibitemShut
  {NoStop}%
\bibitem [{\citenamefont {Lee}\ \emph {et~al.}(2023)\citenamefont {Lee},
  \citenamefont {Mulligan}, \citenamefont {Ringer},\ and\ \citenamefont
  {Yao}}]{Lee:2023urk}%
  \BibitemOpen
  \bibfield  {author} {\bibinfo {author} {\bibfnamefont {Kyle}\ \bibnamefont
  {Lee}}, \bibinfo {author} {\bibfnamefont {James}\ \bibnamefont {Mulligan}},
  \bibinfo {author} {\bibfnamefont {Felix}\ \bibnamefont {Ringer}}, \ and\
  \bibinfo {author} {\bibfnamefont {Xiaojun}\ \bibnamefont {Yao}},\ }\bibfield
  {title} {\enquote {\bibinfo {title} {{Liouvillian dynamics of the open
  Schwinger model: String breaking and kinetic dissipation in a thermal
  medium}},}\ }\href {\doibase 10.1103/PhysRevD.108.094518} {\bibfield
  {journal} {\bibinfo  {journal} {Phys. Rev. D}\ }\textbf {\bibinfo {volume}
  {108}},\ \bibinfo {pages} {094518} (\bibinfo {year} {2023})},\ \Eprint
  {http://arxiv.org/abs/2308.03878} {arXiv:2308.03878 [quant-ph]} \BibitemShut
  {NoStop}%
\bibitem [{\citenamefont {Florio}\ and\ \citenamefont
  {Kharzeev}(2021)}]{Florio:2021xvj}%
  \BibitemOpen
  \bibfield  {author} {\bibinfo {author} {\bibfnamefont {Adrien}\ \bibnamefont
  {Florio}}\ and\ \bibinfo {author} {\bibfnamefont {Dmitri~E.}\ \bibnamefont
  {Kharzeev}},\ }\bibfield  {title} {\enquote {\bibinfo {title} {{Gibbs entropy
  from entanglement in electric quenches}},}\ }\href {\doibase
  10.1103/PhysRevD.104.056021} {\bibfield  {journal} {\bibinfo  {journal}
  {Phys. Rev. D}\ }\textbf {\bibinfo {volume} {104}},\ \bibinfo {pages}
  {056021} (\bibinfo {year} {2021})},\ \Eprint
  {http://arxiv.org/abs/2106.00838} {arXiv:2106.00838 [hep-th]} \BibitemShut
  {NoStop}%
\bibitem [{\citenamefont {Dunne}\ \emph {et~al.}(2023)\citenamefont {Dunne},
  \citenamefont {Florio},\ and\ \citenamefont {Kharzeev}}]{Dunne:2022zlx}%
  \BibitemOpen
  \bibfield  {author} {\bibinfo {author} {\bibfnamefont {Gerald~V.}\
  \bibnamefont {Dunne}}, \bibinfo {author} {\bibfnamefont {Adrien}\
  \bibnamefont {Florio}}, \ and\ \bibinfo {author} {\bibfnamefont {Dmitri~E.}\
  \bibnamefont {Kharzeev}},\ }\bibfield  {title} {\enquote {\bibinfo {title}
  {{Entropy suppression through quantum interference in electric pulses}},}\
  }\href {\doibase 10.1103/PhysRevD.108.L031901} {\bibfield  {journal}
  {\bibinfo  {journal} {Phys. Rev. D}\ }\textbf {\bibinfo {volume} {108}},\
  \bibinfo {pages} {L031901} (\bibinfo {year} {2023})},\ \Eprint
  {http://arxiv.org/abs/2211.13347} {arXiv:2211.13347 [hep-ph]} \BibitemShut
  {NoStop}%
\bibitem [{\citenamefont {Grieninger}\ \emph
  {et~al.}(2023{\natexlab{a}})\citenamefont {Grieninger}, \citenamefont
  {Kharzeev},\ and\ \citenamefont {Zahed}}]{Grieninger:2023ehb}%
  \BibitemOpen
  \bibfield  {author} {\bibinfo {author} {\bibfnamefont {Sebastian}\
  \bibnamefont {Grieninger}}, \bibinfo {author} {\bibfnamefont {Dmitri~E.}\
  \bibnamefont {Kharzeev}}, \ and\ \bibinfo {author} {\bibfnamefont {Ismail}\
  \bibnamefont {Zahed}},\ }\bibfield  {title} {\enquote {\bibinfo {title}
  {{Entanglement in a holographic Schwinger pair with confinement}},}\ }\href
  {\doibase 10.1103/PhysRevD.108.086030} {\bibfield  {journal} {\bibinfo
  {journal} {Phys. Rev. D}\ }\textbf {\bibinfo {volume} {108}},\ \bibinfo
  {pages} {086030} (\bibinfo {year} {2023}{\natexlab{a}})},\ \Eprint
  {http://arxiv.org/abs/2305.07121} {arXiv:2305.07121 [hep-th]} \BibitemShut
  {NoStop}%
\bibitem [{\citenamefont {Grieninger}\ \emph
  {et~al.}(2023{\natexlab{b}})\citenamefont {Grieninger}, \citenamefont
  {Kharzeev},\ and\ \citenamefont {Zahed}}]{Grieninger:2023pyb}%
  \BibitemOpen
  \bibfield  {author} {\bibinfo {author} {\bibfnamefont {Sebastian}\
  \bibnamefont {Grieninger}}, \bibinfo {author} {\bibfnamefont {Dmitri~E.}\
  \bibnamefont {Kharzeev}}, \ and\ \bibinfo {author} {\bibfnamefont {Ismail}\
  \bibnamefont {Zahed}},\ }\bibfield  {title} {\enquote {\bibinfo {title}
  {{Entanglement entropy in a time-dependent holographic Schwinger pair
  creation}},}\ }\href@noop {} {\  (\bibinfo {year} {2023}{\natexlab{b}})},\
  \Eprint {http://arxiv.org/abs/2310.12042} {arXiv:2310.12042 [hep-th]}
  \BibitemShut {NoStop}%
\bibitem [{\citenamefont {Beane}\ \emph {et~al.}(2019)\citenamefont {Beane},
  \citenamefont {Kaplan}, \citenamefont {Klco},\ and\ \citenamefont
  {Savage}}]{Beane:2018oxh}%
  \BibitemOpen
  \bibfield  {author} {\bibinfo {author} {\bibfnamefont {Silas~R.}\
  \bibnamefont {Beane}}, \bibinfo {author} {\bibfnamefont {David~B.}\
  \bibnamefont {Kaplan}}, \bibinfo {author} {\bibfnamefont {Natalie}\
  \bibnamefont {Klco}}, \ and\ \bibinfo {author} {\bibfnamefont {Martin~J.}\
  \bibnamefont {Savage}},\ }\bibfield  {title} {\enquote {\bibinfo {title}
  {{Entanglement Suppression and Emergent Symmetries of Strong
  Interactions}},}\ }\href {\doibase 10.1103/PhysRevLett.122.102001} {\bibfield
   {journal} {\bibinfo  {journal} {Phys. Rev. Lett.}\ }\textbf {\bibinfo
  {volume} {122}},\ \bibinfo {pages} {102001} (\bibinfo {year} {2019})},\
  \Eprint {http://arxiv.org/abs/1812.03138} {arXiv:1812.03138 [nucl-th]}
  \BibitemShut {NoStop}%
\bibitem [{\citenamefont {Klco}\ and\ \citenamefont
  {Savage}(2021{\natexlab{a}})}]{Klco:2020rga}%
  \BibitemOpen
  \bibfield  {author} {\bibinfo {author} {\bibfnamefont {Natalie}\ \bibnamefont
  {Klco}}\ and\ \bibinfo {author} {\bibfnamefont {Martin~J.}\ \bibnamefont
  {Savage}},\ }\bibfield  {title} {\enquote {\bibinfo {title} {{Geometric
  quantum information structure in quantum fields and their lattice
  simulation}},}\ }\href {\doibase 10.1103/PhysRevD.103.065007} {\bibfield
  {journal} {\bibinfo  {journal} {Phys. Rev. D}\ }\textbf {\bibinfo {volume}
  {103}},\ \bibinfo {pages} {065007} (\bibinfo {year} {2021}{\natexlab{a}})},\
  \Eprint {http://arxiv.org/abs/2008.03647} {arXiv:2008.03647 [quant-ph]}
  \BibitemShut {NoStop}%
\bibitem [{\citenamefont {Klco}\ and\ \citenamefont
  {Savage}(2021{\natexlab{b}})}]{Klco:2021biu}%
  \BibitemOpen
  \bibfield  {author} {\bibinfo {author} {\bibfnamefont {Natalie}\ \bibnamefont
  {Klco}}\ and\ \bibinfo {author} {\bibfnamefont {Martin~J.}\ \bibnamefont
  {Savage}},\ }\bibfield  {title} {\enquote {\bibinfo {title} {{Entanglement
  Spheres and a UV-IR Connection in Effective Field Theories}},}\ }\href
  {\doibase 10.1103/PhysRevLett.127.211602} {\bibfield  {journal} {\bibinfo
  {journal} {Phys. Rev. Lett.}\ }\textbf {\bibinfo {volume} {127}},\ \bibinfo
  {pages} {211602} (\bibinfo {year} {2021}{\natexlab{b}})},\ \Eprint
  {http://arxiv.org/abs/2103.14999} {arXiv:2103.14999 [hep-th]} \BibitemShut
  {NoStop}%
\bibitem [{\citenamefont {Klco}\ \emph {et~al.}(2023)\citenamefont {Klco},
  \citenamefont {Beck},\ and\ \citenamefont {Savage}}]{Klco:2021cxq}%
  \BibitemOpen
  \bibfield  {author} {\bibinfo {author} {\bibfnamefont {Natalie}\ \bibnamefont
  {Klco}}, \bibinfo {author} {\bibfnamefont {D.~H.}\ \bibnamefont {Beck}}, \
  and\ \bibinfo {author} {\bibfnamefont {Martin~J.}\ \bibnamefont {Savage}},\
  }\bibfield  {title} {\enquote {\bibinfo {title} {{Entanglement structures in
  quantum field theories: Negativity cores and bound entanglement in the
  vacuum}},}\ }\href {\doibase 10.1103/PhysRevA.107.012415} {\bibfield
  {journal} {\bibinfo  {journal} {Phys. Rev. A}\ }\textbf {\bibinfo {volume}
  {107}},\ \bibinfo {pages} {012415} (\bibinfo {year} {2023})},\ \Eprint
  {http://arxiv.org/abs/2110.10736} {arXiv:2110.10736 [quant-ph]} \BibitemShut
  {NoStop}%
\bibitem [{\citenamefont {Kharzeev}\ and\ \citenamefont
  {Tuchin}(2005)}]{Kharzeev:2005iz}%
  \BibitemOpen
  \bibfield  {author} {\bibinfo {author} {\bibfnamefont {Dmitri}\ \bibnamefont
  {Kharzeev}}\ and\ \bibinfo {author} {\bibfnamefont {Kirill}\ \bibnamefont
  {Tuchin}},\ }\bibfield  {title} {\enquote {\bibinfo {title} {{From color
  glass condensate to quark gluon plasma through the event horizon}},}\ }\href
  {\doibase 10.1016/j.nuclphysa.2005.03.001} {\bibfield  {journal} {\bibinfo
  {journal} {Nucl. Phys. A}\ }\textbf {\bibinfo {volume} {753}},\ \bibinfo
  {pages} {316--334} (\bibinfo {year} {2005})},\ \Eprint
  {http://arxiv.org/abs/hep-ph/0501234} {arXiv:hep-ph/0501234} \BibitemShut
  {NoStop}%
\bibitem [{\citenamefont {Berges}\ \emph {et~al.}(2018)\citenamefont {Berges},
  \citenamefont {Floerchinger},\ and\ \citenamefont
  {Venugopalan}}]{Berges:2017hne}%
  \BibitemOpen
  \bibfield  {author} {\bibinfo {author} {\bibfnamefont {J\"urgen}\
  \bibnamefont {Berges}}, \bibinfo {author} {\bibfnamefont {Stefan}\
  \bibnamefont {Floerchinger}}, \ and\ \bibinfo {author} {\bibfnamefont {Raju}\
  \bibnamefont {Venugopalan}},\ }\bibfield  {title} {\enquote {\bibinfo {title}
  {{Dynamics of entanglement in expanding quantum fields}},}\ }\href {\doibase
  10.1007/JHEP04(2018)145} {\bibfield  {journal} {\bibinfo  {journal} {JHEP}\
  }\textbf {\bibinfo {volume} {04}},\ \bibinfo {pages} {145} (\bibinfo {year}
  {2018})},\ \Eprint {http://arxiv.org/abs/1712.09362} {arXiv:1712.09362
  [hep-th]} \BibitemShut {NoStop}%
\bibitem [{\citenamefont {Zhou}\ \emph {et~al.}(2022)\citenamefont {Zhou},
  \citenamefont {Su}, \citenamefont {Halimeh}, \citenamefont {Ott},
  \citenamefont {Sun}, \citenamefont {Hauke}, \citenamefont {Yang},
  \citenamefont {Yuan}, \citenamefont {Berges},\ and\ \citenamefont
  {Pan}}]{Zhou:2021kdl}%
  \BibitemOpen
  \bibfield  {author} {\bibinfo {author} {\bibfnamefont {Zhao-Yu}\ \bibnamefont
  {Zhou}}, \bibinfo {author} {\bibfnamefont {Guo-Xian}\ \bibnamefont {Su}},
  \bibinfo {author} {\bibfnamefont {Jad~C.}\ \bibnamefont {Halimeh}}, \bibinfo
  {author} {\bibfnamefont {Robert}\ \bibnamefont {Ott}}, \bibinfo {author}
  {\bibfnamefont {Hui}\ \bibnamefont {Sun}}, \bibinfo {author} {\bibfnamefont
  {Philipp}\ \bibnamefont {Hauke}}, \bibinfo {author} {\bibfnamefont {Bing}\
  \bibnamefont {Yang}}, \bibinfo {author} {\bibfnamefont {Zhen-Sheng}\
  \bibnamefont {Yuan}}, \bibinfo {author} {\bibfnamefont {J\"urgen}\
  \bibnamefont {Berges}}, \ and\ \bibinfo {author} {\bibfnamefont {Jian-Wei}\
  \bibnamefont {Pan}},\ }\bibfield  {title} {\enquote {\bibinfo {title}
  {{Thermalization dynamics of a gauge theory on a quantum simulator}},}\
  }\href {\doibase 10.1126/science.abl6277} {\bibfield  {journal} {\bibinfo
  {journal} {Science}\ }\textbf {\bibinfo {volume} {377}},\ \bibinfo {pages}
  {abl6277} (\bibinfo {year} {2022})},\ \Eprint
  {http://arxiv.org/abs/2107.13563} {arXiv:2107.13563 [cond-mat.quant-gas]}
  \BibitemShut {NoStop}%
\bibitem [{\citenamefont {Mueller}\ \emph {et~al.}(2022)\citenamefont
  {Mueller}, \citenamefont {Zache},\ and\ \citenamefont
  {Ott}}]{Mueller:2021gxd}%
  \BibitemOpen
  \bibfield  {author} {\bibinfo {author} {\bibfnamefont {Niklas}\ \bibnamefont
  {Mueller}}, \bibinfo {author} {\bibfnamefont {Torsten~V.}\ \bibnamefont
  {Zache}}, \ and\ \bibinfo {author} {\bibfnamefont {Robert}\ \bibnamefont
  {Ott}},\ }\bibfield  {title} {\enquote {\bibinfo {title} {{Thermalization of
  Gauge Theories from their Entanglement Spectrum}},}\ }\href {\doibase
  10.1103/PhysRevLett.129.011601} {\bibfield  {journal} {\bibinfo  {journal}
  {Phys. Rev. Lett.}\ }\textbf {\bibinfo {volume} {129}},\ \bibinfo {pages}
  {011601} (\bibinfo {year} {2022})},\ \Eprint
  {http://arxiv.org/abs/2107.11416} {arXiv:2107.11416 [quant-ph]} \BibitemShut
  {NoStop}%
\bibitem [{\citenamefont {Ebner}\ \emph {et~al.}(2023)\citenamefont {Ebner},
  \citenamefont {M\"uller}, \citenamefont {Sch\"afer}, \citenamefont {Seidl},\
  and\ \citenamefont {Yao}}]{Ebner:2023ixq}%
  \BibitemOpen
  \bibfield  {author} {\bibinfo {author} {\bibfnamefont {Lukas}\ \bibnamefont
  {Ebner}}, \bibinfo {author} {\bibfnamefont {Berndt}\ \bibnamefont
  {M\"uller}}, \bibinfo {author} {\bibfnamefont {Andreas}\ \bibnamefont
  {Sch\"afer}}, \bibinfo {author} {\bibfnamefont {Clemens}\ \bibnamefont
  {Seidl}}, \ and\ \bibinfo {author} {\bibfnamefont {Xiaojun}\ \bibnamefont
  {Yao}},\ }\bibfield  {title} {\enquote {\bibinfo {title} {{Eigenstate
  Thermalization in 2+1 dimensional SU(2) Lattice Gauge Theory}},}\ }\href@noop
  {} {\  (\bibinfo {year} {2023})},\ \Eprint {http://arxiv.org/abs/2308.16202}
  {arXiv:2308.16202 [hep-lat]} \BibitemShut {NoStop}%
\bibitem [{\citenamefont {Yao}(2023)}]{Yao:2023pht}%
  \BibitemOpen
  \bibfield  {author} {\bibinfo {author} {\bibfnamefont {Xiaojun}\ \bibnamefont
  {Yao}},\ }\bibfield  {title} {\enquote {\bibinfo {title} {{SU(2) gauge theory
  in 2+1 dimensions on a plaquette chain obeys the eigenstate thermalization
  hypothesis}},}\ }\href {\doibase 10.1103/PhysRevD.108.L031504} {\bibfield
  {journal} {\bibinfo  {journal} {Phys. Rev. D}\ }\textbf {\bibinfo {volume}
  {108}},\ \bibinfo {pages} {L031504} (\bibinfo {year} {2023})},\ \Eprint
  {http://arxiv.org/abs/2303.14264} {arXiv:2303.14264 [hep-lat]} \BibitemShut
  {NoStop}%
\bibitem [{\citenamefont {Grieninger}\ \emph
  {et~al.}(2023{\natexlab{c}})\citenamefont {Grieninger}, \citenamefont
  {Ikeda}, \citenamefont {Kharzeev},\ and\ \citenamefont
  {Zahed}}]{Grieninger:2023ufa}%
  \BibitemOpen
  \bibfield  {author} {\bibinfo {author} {\bibfnamefont {Sebastian}\
  \bibnamefont {Grieninger}}, \bibinfo {author} {\bibfnamefont {Kazuki}\
  \bibnamefont {Ikeda}}, \bibinfo {author} {\bibfnamefont {Dmitri~E.}\
  \bibnamefont {Kharzeev}}, \ and\ \bibinfo {author} {\bibfnamefont {Ismail}\
  \bibnamefont {Zahed}},\ }\bibfield  {title} {\enquote {\bibinfo {title}
  {{Entanglement in massive Schwinger model at finite temperature and
  density}},}\ }\href@noop {} {\  (\bibinfo {year} {2023}{\natexlab{c}})},\
  \Eprint {http://arxiv.org/abs/2312.03172} {arXiv:2312.03172 [hep-th]}
  \BibitemShut {NoStop}%
\bibitem [{\citenamefont {Ares}\ \emph
  {et~al.}(2023{\natexlab{a}})\citenamefont {Ares}, \citenamefont {Murciano},\
  and\ \citenamefont {Calabrese}}]{Ares:2022koq}%
  \BibitemOpen
  \bibfield  {author} {\bibinfo {author} {\bibfnamefont {Filiberto}\
  \bibnamefont {Ares}}, \bibinfo {author} {\bibfnamefont {Sara}\ \bibnamefont
  {Murciano}}, \ and\ \bibinfo {author} {\bibfnamefont {Pasquale}\ \bibnamefont
  {Calabrese}},\ }\bibfield  {title} {\enquote {\bibinfo {title} {{Entanglement
  asymmetry as a probe of symmetry breaking}},}\ }\href {\doibase
  10.1038/s41467-023-37747-8} {\bibfield  {journal} {\bibinfo  {journal}
  {Nature Commun.}\ }\textbf {\bibinfo {volume} {14}},\ \bibinfo {pages} {2036}
  (\bibinfo {year} {2023}{\natexlab{a}})},\ \Eprint
  {http://arxiv.org/abs/2207.14693} {arXiv:2207.14693 [cond-mat.stat-mech]}
  \BibitemShut {NoStop}%
\bibitem [{\citenamefont {Ares}\ \emph
  {et~al.}(2023{\natexlab{b}})\citenamefont {Ares}, \citenamefont {Murciano},
  \citenamefont {Vernier},\ and\ \citenamefont {Calabrese}}]{Ares:2023kcz}%
  \BibitemOpen
  \bibfield  {author} {\bibinfo {author} {\bibfnamefont {Filiberto}\
  \bibnamefont {Ares}}, \bibinfo {author} {\bibfnamefont {Sara}\ \bibnamefont
  {Murciano}}, \bibinfo {author} {\bibfnamefont {Eric}\ \bibnamefont
  {Vernier}}, \ and\ \bibinfo {author} {\bibfnamefont {Pasquale}\ \bibnamefont
  {Calabrese}},\ }\bibfield  {title} {\enquote {\bibinfo {title} {{Lack of
  symmetry restoration after a quantum quench: An entanglement asymmetry
  study}},}\ }\href {\doibase 10.21468/SciPostPhys.15.3.089} {\bibfield
  {journal} {\bibinfo  {journal} {SciPost Phys.}\ }\textbf {\bibinfo {volume}
  {15}},\ \bibinfo {pages} {089} (\bibinfo {year} {2023}{\natexlab{b}})},\
  \Eprint {http://arxiv.org/abs/2302.03330} {arXiv:2302.03330
  [cond-mat.stat-mech]} \BibitemShut {NoStop}%
\bibitem [{\citenamefont {Ares}\ \emph
  {et~al.}(2023{\natexlab{c}})\citenamefont {Ares}, \citenamefont {Murciano},
  \citenamefont {Piroli},\ and\ \citenamefont {Calabrese}}]{Ares:2023ggj}%
  \BibitemOpen
  \bibfield  {author} {\bibinfo {author} {\bibfnamefont {Filiberto}\
  \bibnamefont {Ares}}, \bibinfo {author} {\bibfnamefont {Sara}\ \bibnamefont
  {Murciano}}, \bibinfo {author} {\bibfnamefont {Lorenzo}\ \bibnamefont
  {Piroli}}, \ and\ \bibinfo {author} {\bibfnamefont {Pasquale}\ \bibnamefont
  {Calabrese}},\ }\bibfield  {title} {\enquote {\bibinfo {title} {{An
  entanglement asymmetry study of black hole radiation}},}\ }\href@noop {} {\
  (\bibinfo {year} {2023}{\natexlab{c}})},\ \Eprint
  {http://arxiv.org/abs/2311.12683} {arXiv:2311.12683 [hep-th]} \BibitemShut
  {NoStop}%
\bibitem [{\citenamefont {Nishioka}(2018)}]{Nishioka:2018khk}%
  \BibitemOpen
  \bibfield  {author} {\bibinfo {author} {\bibfnamefont {Tatsuma}\ \bibnamefont
  {Nishioka}},\ }\bibfield  {title} {\enquote {\bibinfo {title} {{Entanglement
  entropy: holography and renormalization group}},}\ }\href {\doibase
  10.1103/RevModPhys.90.035007} {\bibfield  {journal} {\bibinfo  {journal}
  {Rev. Mod. Phys.}\ }\textbf {\bibinfo {volume} {90}},\ \bibinfo {pages}
  {035007} (\bibinfo {year} {2018})},\ \Eprint
  {http://arxiv.org/abs/1801.10352} {arXiv:1801.10352 [hep-th]} \BibitemShut
  {NoStop}%
\bibitem [{\citenamefont {Klco}\ \emph {et~al.}(2022)\citenamefont {Klco},
  \citenamefont {Roggero},\ and\ \citenamefont {Savage}}]{Klco:2021lap}%
  \BibitemOpen
  \bibfield  {author} {\bibinfo {author} {\bibfnamefont {Natalie}\ \bibnamefont
  {Klco}}, \bibinfo {author} {\bibfnamefont {Alessandro}\ \bibnamefont
  {Roggero}}, \ and\ \bibinfo {author} {\bibfnamefont {Martin~J.}\ \bibnamefont
  {Savage}},\ }\bibfield  {title} {\enquote {\bibinfo {title} {{Standard model
  physics and the digital quantum revolution: thoughts about the interface}},}\
  }\href {\doibase 10.1088/1361-6633/ac58a4} {\bibfield  {journal} {\bibinfo
  {journal} {Rept. Prog. Phys.}\ }\textbf {\bibinfo {volume} {85}},\ \bibinfo
  {pages} {064301} (\bibinfo {year} {2022})},\ \Eprint
  {http://arxiv.org/abs/2107.04769} {arXiv:2107.04769 [quant-ph]} \BibitemShut
  {NoStop}%
\bibitem [{\citenamefont {Bauer}\ \emph {et~al.}(2023)\citenamefont {Bauer},
  \citenamefont {Davoudi}, \citenamefont {Klco},\ and\ \citenamefont
  {Savage}}]{Bauer:2023qgm}%
  \BibitemOpen
  \bibfield  {author} {\bibinfo {author} {\bibfnamefont {Christian~W.}\
  \bibnamefont {Bauer}}, \bibinfo {author} {\bibfnamefont {Zohreh}\
  \bibnamefont {Davoudi}}, \bibinfo {author} {\bibfnamefont {Natalie}\
  \bibnamefont {Klco}}, \ and\ \bibinfo {author} {\bibfnamefont {Martin~J.}\
  \bibnamefont {Savage}},\ }\bibfield  {title} {\enquote {\bibinfo {title}
  {{Quantum simulation of fundamental particles and forces}},}\ }\href
  {\doibase 10.1038/s42254-023-00599-8} {\bibfield  {journal} {\bibinfo
  {journal} {Nature Rev. Phys.}\ }\textbf {\bibinfo {volume} {5}},\ \bibinfo
  {pages} {420--432} (\bibinfo {year} {2023})}\BibitemShut {NoStop}%
\bibitem [{\citenamefont {Peres}(1996)}]{Peres:1996dw}%
  \BibitemOpen
  \bibfield  {author} {\bibinfo {author} {\bibfnamefont {Asher}\ \bibnamefont
  {Peres}},\ }\bibfield  {title} {\enquote {\bibinfo {title} {{Separability
  criterion for density matrices}},}\ }\href {\doibase
  10.1103/PhysRevLett.77.1413} {\bibfield  {journal} {\bibinfo  {journal}
  {Phys. Rev. Lett.}\ }\textbf {\bibinfo {volume} {77}},\ \bibinfo {pages}
  {1413--1415} (\bibinfo {year} {1996})},\ \Eprint
  {http://arxiv.org/abs/quant-ph/9604005} {arXiv:quant-ph/9604005} \BibitemShut
  {NoStop}%
\bibitem [{\citenamefont {Horodecki}(1997)}]{Horodecki:1997vt}%
  \BibitemOpen
  \bibfield  {author} {\bibinfo {author} {\bibfnamefont {Pawel}\ \bibnamefont
  {Horodecki}},\ }\bibfield  {title} {\enquote {\bibinfo {title} {{Separability
  criterion and inseparable mixed states with positive partial
  transposition}},}\ }\href {\doibase 10.1016/S0375-9601(97)00416-7} {\bibfield
   {journal} {\bibinfo  {journal} {Phys. Lett. A}\ }\textbf {\bibinfo {volume}
  {232}},\ \bibinfo {pages} {333} (\bibinfo {year} {1997})},\ \Eprint
  {http://arxiv.org/abs/quant-ph/9703004} {arXiv:quant-ph/9703004} \BibitemShut
  {NoStop}%
\bibitem [{\citenamefont {Simon}(2000)}]{Simon:1999lfr}%
  \BibitemOpen
  \bibfield  {author} {\bibinfo {author} {\bibfnamefont {R.}~\bibnamefont
  {Simon}},\ }\bibfield  {title} {\enquote {\bibinfo {title} {{Peres-Horodecki
  Separability Criterion for Continuous Variable Systems}},}\ }\href {\doibase
  10.1103/PhysRevLett.84.2726} {\bibfield  {journal} {\bibinfo  {journal}
  {Phys. Rev. Lett.}\ }\textbf {\bibinfo {volume} {84}},\ \bibinfo {pages}
  {2726--2729} (\bibinfo {year} {2000})},\ \Eprint
  {http://arxiv.org/abs/quant-ph/9909044} {arXiv:quant-ph/9909044} \BibitemShut
  {NoStop}%
\bibitem [{\citenamefont {Vidal}\ and\ \citenamefont
  {Werner}(2002)}]{Vidal:2002zz}%
  \BibitemOpen
  \bibfield  {author} {\bibinfo {author} {\bibfnamefont {G.}~\bibnamefont
  {Vidal}}\ and\ \bibinfo {author} {\bibfnamefont {R.~F.}\ \bibnamefont
  {Werner}},\ }\bibfield  {title} {\enquote {\bibinfo {title} {{Computable
  measure of entanglement}},}\ }\href {\doibase 10.1103/PhysRevA.65.032314}
  {\bibfield  {journal} {\bibinfo  {journal} {Phys. Rev. A}\ }\textbf {\bibinfo
  {volume} {65}},\ \bibinfo {pages} {032314} (\bibinfo {year} {2002})},\
  \Eprint {http://arxiv.org/abs/quant-ph/0102117} {arXiv:quant-ph/0102117}
  \BibitemShut {NoStop}%
\bibitem [{\citenamefont {Plenio}(2005)}]{Plenio:2005cwa}%
  \BibitemOpen
  \bibfield  {author} {\bibinfo {author} {\bibfnamefont {M.~B.}\ \bibnamefont
  {Plenio}},\ }\bibfield  {title} {\enquote {\bibinfo {title} {{Logarithmic
  Negativity: A Full Entanglement Monotone That is not Convex}},}\ }\href
  {\doibase 10.1103/PhysRevLett.95.090503} {\bibfield  {journal} {\bibinfo
  {journal} {Phys. Rev. Lett.}\ }\textbf {\bibinfo {volume} {95}},\ \bibinfo
  {pages} {090503} (\bibinfo {year} {2005})},\ \Eprint
  {http://arxiv.org/abs/quant-ph/0505071} {arXiv:quant-ph/0505071} \BibitemShut
  {NoStop}%
\bibitem [{\citenamefont {Calabrese}\ \emph {et~al.}(2012)\citenamefont
  {Calabrese}, \citenamefont {Cardy},\ and\ \citenamefont
  {Tonni}}]{Calabrese:2012ew}%
  \BibitemOpen
  \bibfield  {author} {\bibinfo {author} {\bibfnamefont {Pasquale}\
  \bibnamefont {Calabrese}}, \bibinfo {author} {\bibfnamefont {John}\
  \bibnamefont {Cardy}}, \ and\ \bibinfo {author} {\bibfnamefont {Erik}\
  \bibnamefont {Tonni}},\ }\bibfield  {title} {\enquote {\bibinfo {title}
  {{Entanglement negativity in quantum field theory}},}\ }\href {\doibase
  10.1103/PhysRevLett.109.130502} {\bibfield  {journal} {\bibinfo  {journal}
  {Phys. Rev. Lett.}\ }\textbf {\bibinfo {volume} {109}},\ \bibinfo {pages}
  {130502} (\bibinfo {year} {2012})},\ \Eprint {http://arxiv.org/abs/1206.3092}
  {arXiv:1206.3092 [cond-mat.stat-mech]} \BibitemShut {NoStop}%
\bibitem [{\citenamefont {Calabrese}\ \emph {et~al.}(2013)\citenamefont
  {Calabrese}, \citenamefont {Cardy},\ and\ \citenamefont
  {Tonni}}]{Calabrese:2012nk}%
  \BibitemOpen
  \bibfield  {author} {\bibinfo {author} {\bibfnamefont {Pasquale}\
  \bibnamefont {Calabrese}}, \bibinfo {author} {\bibfnamefont {John}\
  \bibnamefont {Cardy}}, \ and\ \bibinfo {author} {\bibfnamefont {Erik}\
  \bibnamefont {Tonni}},\ }\bibfield  {title} {\enquote {\bibinfo {title}
  {{Entanglement negativity in extended systems: A field theoretical
  approach}},}\ }\href {\doibase 10.1088/1742-5468/2013/02/P02008} {\bibfield
  {journal} {\bibinfo  {journal} {J. Stat. Mech.}\ }\textbf {\bibinfo {volume}
  {1302}},\ \bibinfo {pages} {P02008} (\bibinfo {year} {2013})},\ \Eprint
  {http://arxiv.org/abs/1210.5359} {arXiv:1210.5359 [cond-mat.stat-mech]}
  \BibitemShut {NoStop}%
\bibitem [{\citenamefont {Murciano}\ \emph {et~al.}(2021)\citenamefont
  {Murciano}, \citenamefont {Bonsignori},\ and\ \citenamefont
  {Calabrese}}]{Murciano:2021djk}%
  \BibitemOpen
  \bibfield  {author} {\bibinfo {author} {\bibfnamefont {Sara}\ \bibnamefont
  {Murciano}}, \bibinfo {author} {\bibfnamefont {Riccarda}\ \bibnamefont
  {Bonsignori}}, \ and\ \bibinfo {author} {\bibfnamefont {Pasquale}\
  \bibnamefont {Calabrese}},\ }\bibfield  {title} {\enquote {\bibinfo {title}
  {{Symmetry decomposition of negativity of massless free fermions}},}\ }\href
  {\doibase 10.21468/SciPostPhys.10.5.111} {\bibfield  {journal} {\bibinfo
  {journal} {SciPost Phys.}\ }\textbf {\bibinfo {volume} {10}},\ \bibinfo
  {pages} {111} (\bibinfo {year} {2021})},\ \Eprint
  {http://arxiv.org/abs/2102.10054} {arXiv:2102.10054 [cond-mat.stat-mech]}
  \BibitemShut {NoStop}%
\bibitem [{\citenamefont {Murciano}\ \emph {et~al.}(2022)\citenamefont
  {Murciano}, \citenamefont {Vitale}, \citenamefont {Dalmonte},\ and\
  \citenamefont {Calabrese}}]{Murciano:2022vhe}%
  \BibitemOpen
  \bibfield  {author} {\bibinfo {author} {\bibfnamefont {Sara}\ \bibnamefont
  {Murciano}}, \bibinfo {author} {\bibfnamefont {Vittorio}\ \bibnamefont
  {Vitale}}, \bibinfo {author} {\bibfnamefont {Marcello}\ \bibnamefont
  {Dalmonte}}, \ and\ \bibinfo {author} {\bibfnamefont {Pasquale}\ \bibnamefont
  {Calabrese}},\ }\bibfield  {title} {\enquote {\bibinfo {title} {{Negativity
  Hamiltonian: An Operator Characterization of Mixed-State Entanglement}},}\
  }\href {\doibase 10.1103/PhysRevLett.128.140502} {\bibfield  {journal}
  {\bibinfo  {journal} {Phys. Rev. Lett.}\ }\textbf {\bibinfo {volume} {128}},\
  \bibinfo {pages} {140502} (\bibinfo {year} {2022})},\ \Eprint
  {http://arxiv.org/abs/2201.03989} {arXiv:2201.03989 [cond-mat.stat-mech]}
  \BibitemShut {NoStop}%
\bibitem [{\citenamefont {Parez}\ and\ \citenamefont
  {Witczak-Krempa}(2023)}]{Parez:2023uxu}%
  \BibitemOpen
  \bibfield  {author} {\bibinfo {author} {\bibfnamefont {Gilles}\ \bibnamefont
  {Parez}}\ and\ \bibinfo {author} {\bibfnamefont {William}\ \bibnamefont
  {Witczak-Krempa}},\ }\bibfield  {title} {\enquote {\bibinfo {title} {{Are
  fermionic conformal field theories more entangled?}}}\ }\href@noop {} {\
  (\bibinfo {year} {2023})},\ \Eprint {http://arxiv.org/abs/2310.15273}
  {arXiv:2310.15273 [cond-mat.str-el]} \BibitemShut {NoStop}%
\bibitem [{\citenamefont {Marcovitch}\ \emph {et~al.}(2009)\citenamefont
  {Marcovitch}, \citenamefont {Retzker}, \citenamefont {Plenio},\ and\
  \citenamefont {Reznik}}]{Marcovitch:2008sxc}%
  \BibitemOpen
  \bibfield  {author} {\bibinfo {author} {\bibfnamefont {S.}~\bibnamefont
  {Marcovitch}}, \bibinfo {author} {\bibfnamefont {A.}~\bibnamefont {Retzker}},
  \bibinfo {author} {\bibfnamefont {M.~B.}\ \bibnamefont {Plenio}}, \ and\
  \bibinfo {author} {\bibfnamefont {B.}~\bibnamefont {Reznik}},\ }\bibfield
  {title} {\enquote {\bibinfo {title} {{Critical and noncritical long-range
  entanglement in Klein-Gordon fields}},}\ }\href {\doibase
  10.1103/PhysRevA.80.012325} {\bibfield  {journal} {\bibinfo  {journal} {Phys.
  Rev. A}\ }\textbf {\bibinfo {volume} {80}},\ \bibinfo {pages} {012325}
  (\bibinfo {year} {2009})},\ \Eprint {http://arxiv.org/abs/0811.1288}
  {arXiv:0811.1288 [quant-ph]} \BibitemShut {NoStop}%
\bibitem [{\citenamefont {Coleman}\ \emph {et~al.}(1975)\citenamefont
  {Coleman}, \citenamefont {Jackiw},\ and\ \citenamefont
  {Susskind}}]{Coleman:1975pw}%
  \BibitemOpen
  \bibfield  {author} {\bibinfo {author} {\bibfnamefont {Sidney~R.}\
  \bibnamefont {Coleman}}, \bibinfo {author} {\bibfnamefont {R.}~\bibnamefont
  {Jackiw}}, \ and\ \bibinfo {author} {\bibfnamefont {Leonard}\ \bibnamefont
  {Susskind}},\ }\bibfield  {title} {\enquote {\bibinfo {title} {{Charge
  Shielding and Quark Confinement in the Massive Schwinger Model}},}\ }\href
  {\doibase 10.1016/0003-4916(75)90212-2} {\bibfield  {journal} {\bibinfo
  {journal} {Annals Phys.}\ }\textbf {\bibinfo {volume} {93}},\ \bibinfo
  {pages} {267} (\bibinfo {year} {1975})}\BibitemShut {NoStop}%
\bibitem [{\citenamefont {Coleman}(1976)}]{Coleman:1976uz}%
  \BibitemOpen
  \bibfield  {author} {\bibinfo {author} {\bibfnamefont {Sidney~R.}\
  \bibnamefont {Coleman}},\ }\bibfield  {title} {\enquote {\bibinfo {title}
  {{More About the Massive Schwinger Model}},}\ }\href {\doibase
  10.1016/0003-4916(76)90280-3} {\bibfield  {journal} {\bibinfo  {journal}
  {Annals Phys.}\ }\textbf {\bibinfo {volume} {101}},\ \bibinfo {pages} {239}
  (\bibinfo {year} {1976})}\BibitemShut {NoStop}%
\bibitem [{\citenamefont {Schwinger}(1962)}]{Schwinger:1962tp}%
  \BibitemOpen
  \bibfield  {author} {\bibinfo {author} {\bibfnamefont {Julian~S.}\
  \bibnamefont {Schwinger}},\ }\bibfield  {title} {\enquote {\bibinfo {title}
  {{Gauge Invariance and Mass. 2.}}}\ }\href {\doibase
  10.1103/PhysRev.128.2425} {\bibfield  {journal} {\bibinfo  {journal} {Phys.
  Rev.}\ }\textbf {\bibinfo {volume} {128}},\ \bibinfo {pages} {2425--2429}
  (\bibinfo {year} {1962})}\BibitemShut {NoStop}%
\bibitem [{\citenamefont {Kogut}\ and\ \citenamefont
  {Susskind}(1975)}]{Kogut:1974ag}%
  \BibitemOpen
  \bibfield  {author} {\bibinfo {author} {\bibfnamefont {John~B.}\ \bibnamefont
  {Kogut}}\ and\ \bibinfo {author} {\bibfnamefont {Leonard}\ \bibnamefont
  {Susskind}},\ }\bibfield  {title} {\enquote {\bibinfo {title} {{Hamiltonian
  Formulation of Wilson's Lattice Gauge Theories}},}\ }\href {\doibase
  10.1103/PhysRevD.11.395} {\bibfield  {journal} {\bibinfo  {journal} {Phys.
  Rev. D}\ }\textbf {\bibinfo {volume} {11}},\ \bibinfo {pages} {395--408}
  (\bibinfo {year} {1975})}\BibitemShut {NoStop}%
\bibitem [{\citenamefont {Susskind}(1977)}]{Susskind:1976jm}%
  \BibitemOpen
  \bibfield  {author} {\bibinfo {author} {\bibfnamefont {Leonard}\ \bibnamefont
  {Susskind}},\ }\bibfield  {title} {\enquote {\bibinfo {title} {{Lattice
  Fermions}},}\ }\href {\doibase 10.1103/PhysRevD.16.3031} {\bibfield
  {journal} {\bibinfo  {journal} {Phys. Rev. D}\ }\textbf {\bibinfo {volume}
  {16}},\ \bibinfo {pages} {3031--3039} (\bibinfo {year} {1977})}\BibitemShut
  {NoStop}%
\bibitem [{\citenamefont {Ikeda}\ \emph {et~al.}(2021)\citenamefont {Ikeda},
  \citenamefont {Kharzeev},\ and\ \citenamefont {Kikuchi}}]{Ikeda:2020agk}%
  \BibitemOpen
  \bibfield  {author} {\bibinfo {author} {\bibfnamefont {Kazuki}\ \bibnamefont
  {Ikeda}}, \bibinfo {author} {\bibfnamefont {Dmitri~E.}\ \bibnamefont
  {Kharzeev}}, \ and\ \bibinfo {author} {\bibfnamefont {Yuta}\ \bibnamefont
  {Kikuchi}},\ }\bibfield  {title} {\enquote {\bibinfo {title} {{Real-time
  dynamics of Chern-Simons fluctuations near a critical point}},}\ }\href
  {\doibase 10.1103/PhysRevD.103.L071502} {\bibfield  {journal} {\bibinfo
  {journal} {Phys. Rev. D}\ }\textbf {\bibinfo {volume} {103}},\ \bibinfo
  {pages} {L071502} (\bibinfo {year} {2021})},\ \Eprint
  {http://arxiv.org/abs/2012.02926} {arXiv:2012.02926 [hep-ph]} \BibitemShut
  {NoStop}%
\bibitem [{\citenamefont {Javanmard}\ \emph {et~al.}(2018)\citenamefont
  {Javanmard}, \citenamefont {Trapin}, \citenamefont {Bera}, \citenamefont
  {Bardarson},\ and\ \citenamefont {Heyl}}]{Javanmard:2018xte}%
  \BibitemOpen
  \bibfield  {author} {\bibinfo {author} {\bibfnamefont {Younes}\ \bibnamefont
  {Javanmard}}, \bibinfo {author} {\bibfnamefont {Daniele}\ \bibnamefont
  {Trapin}}, \bibinfo {author} {\bibfnamefont {Soumya}\ \bibnamefont {Bera}},
  \bibinfo {author} {\bibfnamefont {Jens~H.}\ \bibnamefont {Bardarson}}, \ and\
  \bibinfo {author} {\bibfnamefont {Markus}\ \bibnamefont {Heyl}},\ }\bibfield
  {title} {\enquote {\bibinfo {title} {{Sharp entanglement thresholds in the
  logarithmic negativity of disjoint blocks in the transverse-field Ising
  chain}},}\ }\href {\doibase 10.1088/1367-2630/aad9ba} {\bibfield  {journal}
  {\bibinfo  {journal} {New J. Phys.}\ }\textbf {\bibinfo {volume} {20}},\
  \bibinfo {pages} {083032} (\bibinfo {year} {2018})},\ \Eprint
  {http://arxiv.org/abs/1805.06935} {arXiv:1805.06935 [cond-mat.str-el]}
  \BibitemShut {NoStop}%
\bibitem [{\citenamefont {Berthiere}\ and\ \citenamefont
  {Witczak-Krempa}(2022)}]{Berthiere:2021nkv}%
  \BibitemOpen
  \bibfield  {author} {\bibinfo {author} {\bibfnamefont {Cl\'ement}\
  \bibnamefont {Berthiere}}\ and\ \bibinfo {author} {\bibfnamefont {William}\
  \bibnamefont {Witczak-Krempa}},\ }\bibfield  {title} {\enquote {\bibinfo
  {title} {{Entanglement of Skeletal Regions}},}\ }\href {\doibase
  10.1103/PhysRevLett.128.240502} {\bibfield  {journal} {\bibinfo  {journal}
  {Phys. Rev. Lett.}\ }\textbf {\bibinfo {volume} {128}},\ \bibinfo {pages}
  {240502} (\bibinfo {year} {2022})},\ \Eprint
  {http://arxiv.org/abs/2112.13931} {arXiv:2112.13931 [cond-mat.str-el]}
  \BibitemShut {NoStop}%
\bibitem [{\citenamefont {Shapourian}\ \emph {et~al.}(2017)\citenamefont
  {Shapourian}, \citenamefont {Shiozaki},\ and\ \citenamefont
  {Ryu}}]{Shapourian:2016cqu}%
  \BibitemOpen
  \bibfield  {author} {\bibinfo {author} {\bibfnamefont {Hassan}\ \bibnamefont
  {Shapourian}}, \bibinfo {author} {\bibfnamefont {Ken}\ \bibnamefont
  {Shiozaki}}, \ and\ \bibinfo {author} {\bibfnamefont {Shinsei}\ \bibnamefont
  {Ryu}},\ }\bibfield  {title} {\enquote {\bibinfo {title} {{Partial
  time-reversal transformation and entanglement negativity in fermionic
  systems}},}\ }\href {\doibase 10.1103/PhysRevB.95.165101} {\bibfield
  {journal} {\bibinfo  {journal} {Phys. Rev. B}\ }\textbf {\bibinfo {volume}
  {95}},\ \bibinfo {pages} {165101} (\bibinfo {year} {2017})},\ \Eprint
  {http://arxiv.org/abs/1611.07536} {arXiv:1611.07536 [cond-mat.str-el]}
  \BibitemShut {NoStop}%
\bibitem [{\citenamefont {Malla}\ \emph {et~al.}(2023)\citenamefont {Malla},
  \citenamefont {Weichselbaum}, \citenamefont {Wei},\ and\ \citenamefont
  {Konik}}]{Malla:2023cvl}%
  \BibitemOpen
  \bibfield  {author} {\bibinfo {author} {\bibfnamefont {Rajesh~K.}\
  \bibnamefont {Malla}}, \bibinfo {author} {\bibfnamefont {Andreas}\
  \bibnamefont {Weichselbaum}}, \bibinfo {author} {\bibfnamefont {Tzu-Chieh}\
  \bibnamefont {Wei}}, \ and\ \bibinfo {author} {\bibfnamefont {Robert~M.}\
  \bibnamefont {Konik}},\ }\bibfield  {title} {\enquote {\bibinfo {title}
  {{Detecting Multipartite Entanglement Patterns using Single Particle Green's
  Functions}},}\ }\href@noop {} {\  (\bibinfo {year} {2023})},\ \Eprint
  {http://arxiv.org/abs/2310.05870} {arXiv:2310.05870 [quant-ph]} \BibitemShut
  {NoStop}%
\bibitem [{\citenamefont {Casher}\ \emph {et~al.}(1973)\citenamefont {Casher},
  \citenamefont {Kogut},\ and\ \citenamefont {Susskind}}]{Casher:1973uf}%
  \BibitemOpen
  \bibfield  {author} {\bibinfo {author} {\bibfnamefont {A.}~\bibnamefont
  {Casher}}, \bibinfo {author} {\bibfnamefont {John~B.}\ \bibnamefont {Kogut}},
  \ and\ \bibinfo {author} {\bibfnamefont {Leonard}\ \bibnamefont {Susskind}},\
  }\bibfield  {title} {\enquote {\bibinfo {title} {{Vacuum polarization and the
  quark parton puzzle}},}\ }\href {\doibase 10.1103/PhysRevLett.31.792}
  {\bibfield  {journal} {\bibinfo  {journal} {Phys. Rev. Lett.}\ }\textbf
  {\bibinfo {volume} {31}},\ \bibinfo {pages} {792--795} (\bibinfo {year}
  {1973})}\BibitemShut {NoStop}%
\bibitem [{\citenamefont {Steele}\ \emph {et~al.}(1995)\citenamefont {Steele},
  \citenamefont {Verbaarschot},\ and\ \citenamefont {Zahed}}]{Steele:1994gf}%
  \BibitemOpen
  \bibfield  {author} {\bibinfo {author} {\bibfnamefont {James~V.}\
  \bibnamefont {Steele}}, \bibinfo {author} {\bibfnamefont {J.~J.~M.}\
  \bibnamefont {Verbaarschot}}, \ and\ \bibinfo {author} {\bibfnamefont
  {I.}~\bibnamefont {Zahed}},\ }\bibfield  {title} {\enquote {\bibinfo {title}
  {{The Invariant fermion correlator in the Schwinger model on the torus}},}\
  }\href {\doibase 10.1103/PhysRevD.51.5915} {\bibfield  {journal} {\bibinfo
  {journal} {Phys. Rev. D}\ }\textbf {\bibinfo {volume} {51}},\ \bibinfo
  {pages} {5915--5923} (\bibinfo {year} {1995})},\ \Eprint
  {http://arxiv.org/abs/hep-th/9407125} {arXiv:hep-th/9407125} \BibitemShut
  {NoStop}%
\bibitem [{\citenamefont {Fishman}\ \emph
  {et~al.}(2022{\natexlab{a}})\citenamefont {Fishman}, \citenamefont {White},\
  and\ \citenamefont {Stoudenmire}}]{itensor}%
  \BibitemOpen
  \bibfield  {author} {\bibinfo {author} {\bibfnamefont {Matthew}\ \bibnamefont
  {Fishman}}, \bibinfo {author} {\bibfnamefont {Steven~R.}\ \bibnamefont
  {White}}, \ and\ \bibinfo {author} {\bibfnamefont {E.~Miles}\ \bibnamefont
  {Stoudenmire}},\ }\bibfield  {title} {\enquote {\bibinfo {title} {{The
  ITensor Software Library for Tensor Network Calculations}},}\ }\href
  {\doibase 10.21468/SciPostPhysCodeb.4} {\bibfield  {journal} {\bibinfo
  {journal} {SciPost Phys. Codebases}\ ,\ \bibinfo {pages} {4}} (\bibinfo
  {year} {2022}{\natexlab{a}})}\BibitemShut {NoStop}%
\bibitem [{\citenamefont {Fishman}\ \emph
  {et~al.}(2022{\natexlab{b}})\citenamefont {Fishman}, \citenamefont {White},\
  and\ \citenamefont {Stoudenmire}}]{itensor-r0.3}%
  \BibitemOpen
  \bibfield  {author} {\bibinfo {author} {\bibfnamefont {Matthew}\ \bibnamefont
  {Fishman}}, \bibinfo {author} {\bibfnamefont {Steven~R.}\ \bibnamefont
  {White}}, \ and\ \bibinfo {author} {\bibfnamefont {E.~Miles}\ \bibnamefont
  {Stoudenmire}},\ }\bibfield  {title} {\enquote {\bibinfo {title} {{Codebase
  release 0.3 for ITensor}},}\ }\href {\doibase
  10.21468/SciPostPhysCodeb.4-r0.3} {\bibfield  {journal} {\bibinfo  {journal}
  {SciPost Phys. Codebases}\ ,\ \bibinfo {pages} {4--r0.3}} (\bibinfo {year}
  {2022}{\natexlab{b}})}\BibitemShut {NoStop}%
\bibitem [{\citenamefont {Dempsey}\ \emph {et~al.}(2022)\citenamefont
  {Dempsey}, \citenamefont {Klebanov}, \citenamefont {Pufu},\ and\
  \citenamefont {Zan}}]{Dempsey:2022nys}%
  \BibitemOpen
  \bibfield  {author} {\bibinfo {author} {\bibfnamefont {Ross}\ \bibnamefont
  {Dempsey}}, \bibinfo {author} {\bibfnamefont {Igor~R.}\ \bibnamefont
  {Klebanov}}, \bibinfo {author} {\bibfnamefont {Silviu~S.}\ \bibnamefont
  {Pufu}}, \ and\ \bibinfo {author} {\bibfnamefont {Bernardo}\ \bibnamefont
  {Zan}},\ }\bibfield  {title} {\enquote {\bibinfo {title} {{Discrete chiral
  symmetry and mass shift in the lattice Hamiltonian approach to the Schwinger
  model}},}\ }\href {\doibase 10.1103/PhysRevResearch.4.043133} {\bibfield
  {journal} {\bibinfo  {journal} {Phys. Rev. Res.}\ }\textbf {\bibinfo {volume}
  {4}},\ \bibinfo {pages} {043133} (\bibinfo {year} {2022})},\ \Eprint
  {http://arxiv.org/abs/2206.05308} {arXiv:2206.05308 [hep-th]} \BibitemShut
  {NoStop}%
\bibitem [{\citenamefont {Adam}(2003)}]{Adam:2002pd}%
  \BibitemOpen
  \bibfield  {author} {\bibinfo {author} {\bibfnamefont {C.}~\bibnamefont
  {Adam}},\ }\bibfield  {title} {\enquote {\bibinfo {title} {{Improved vector
  and scalar masses in the massive Schwinger model}},}\ }\href {\doibase
  10.1016/S0370-2693(03)00051-0} {\bibfield  {journal} {\bibinfo  {journal}
  {Phys. Lett. B}\ }\textbf {\bibinfo {volume} {555}},\ \bibinfo {pages}
  {132--137} (\bibinfo {year} {2003})},\ \Eprint
  {http://arxiv.org/abs/hep-th/0212171} {arXiv:hep-th/0212171} \BibitemShut
  {NoStop}%
\bibitem [{\citenamefont {Byrnes}\ \emph
  {et~al.}(2002{\natexlab{a}})\citenamefont {Byrnes}, \citenamefont
  {Sriganesh}, \citenamefont {Bursill},\ and\ \citenamefont
  {Hamer}}]{Byrnes:2002gj}%
  \BibitemOpen
  \bibfield  {author} {\bibinfo {author} {\bibfnamefont {T.}~\bibnamefont
  {Byrnes}}, \bibinfo {author} {\bibfnamefont {P.}~\bibnamefont {Sriganesh}},
  \bibinfo {author} {\bibfnamefont {R.~J.}\ \bibnamefont {Bursill}}, \ and\
  \bibinfo {author} {\bibfnamefont {C.~J.}\ \bibnamefont {Hamer}},\ }\bibfield
  {title} {\enquote {\bibinfo {title} {{Density matrix renormalization group
  approach to the massive Schwinger model}},}\ }\href {\doibase
  10.1016/S0920-5632(02)01416-0} {\bibfield  {journal} {\bibinfo  {journal}
  {Nucl. Phys. B Proc. Suppl.}\ }\textbf {\bibinfo {volume} {109}},\ \bibinfo
  {pages} {202--206} (\bibinfo {year} {2002}{\natexlab{a}})},\ \Eprint
  {http://arxiv.org/abs/hep-lat/0201007} {arXiv:hep-lat/0201007} \BibitemShut
  {NoStop}%
\bibitem [{\citenamefont {Byrnes}\ \emph
  {et~al.}(2002{\natexlab{b}})\citenamefont {Byrnes}, \citenamefont
  {Sriganesh}, \citenamefont {Bursill},\ and\ \citenamefont
  {Hamer}}]{Byrnes:2002nv}%
  \BibitemOpen
  \bibfield  {author} {\bibinfo {author} {\bibfnamefont {T.}~\bibnamefont
  {Byrnes}}, \bibinfo {author} {\bibfnamefont {P.}~\bibnamefont {Sriganesh}},
  \bibinfo {author} {\bibfnamefont {R.~J.}\ \bibnamefont {Bursill}}, \ and\
  \bibinfo {author} {\bibfnamefont {C.~J.}\ \bibnamefont {Hamer}},\ }\bibfield
  {title} {\enquote {\bibinfo {title} {{Density matrix renormalization group
  approach to the massive Schwinger model}},}\ }\href {\doibase
  10.1103/PhysRevD.66.013002} {\bibfield  {journal} {\bibinfo  {journal} {Phys.
  Rev. D}\ }\textbf {\bibinfo {volume} {66}},\ \bibinfo {pages} {013002}
  (\bibinfo {year} {2002}{\natexlab{b}})},\ \Eprint
  {http://arxiv.org/abs/hep-lat/0202014} {arXiv:hep-lat/0202014} \BibitemShut
  {NoStop}%
\bibitem [{\citenamefont {Vidal}(2009)}]{Vidal:2009cbi}%
  \BibitemOpen
  \bibfield  {author} {\bibinfo {author} {\bibfnamefont {Guifre}\ \bibnamefont
  {Vidal}},\ }\bibfield  {title} {\enquote {\bibinfo {title} {{Entanglement
  Renormalization: an introduction}},}\ }\href@noop {} {\  (\bibinfo {year}
  {2009})},\ \Eprint {http://arxiv.org/abs/0912.1651} {arXiv:0912.1651
  [cond-mat.str-el]} \BibitemShut {NoStop}%
\bibitem [{\citenamefont {Bridgeman}\ and\ \citenamefont
  {Chubb}(2017)}]{Bridgeman:2016dhh}%
  \BibitemOpen
  \bibfield  {author} {\bibinfo {author} {\bibfnamefont {Jacob~C.}\
  \bibnamefont {Bridgeman}}\ and\ \bibinfo {author} {\bibfnamefont
  {Christopher~T.}\ \bibnamefont {Chubb}},\ }\bibfield  {title} {\enquote
  {\bibinfo {title} {{Hand-waving and Interpretive Dance: An Introductory
  Course on Tensor Networks}},}\ }\href {\doibase 10.1088/1751-8121/aa6dc3}
  {\bibfield  {journal} {\bibinfo  {journal} {J. Phys. A}\ }\textbf {\bibinfo
  {volume} {50}},\ \bibinfo {pages} {223001} (\bibinfo {year} {2017})},\
  \Eprint {http://arxiv.org/abs/1603.03039} {arXiv:1603.03039 [quant-ph]}
  \BibitemShut {NoStop}%
\end{thebibliography}
%merlin.mbs apsrev4-1.bst 2010-07-25 4.21a (PWD, AO, DPC) hacked
%Control: key (0)
%Control: author (0) dotless jnrlst
%Control: editor formatted (1) identically to author
%Control: production of article title (0) allowed
%Control: page (1) range
%Control: year (0) verbatim
%Control: production of eprint (0) enabled
%

\clearpage

\end{document}